# An Evening Spent with Bill van Zwet

**R. J. Beran and N. I. Fisher**

*Abstract.* Willem Rutger van Zwet was born in Leiden, the Netherlands, on March 31, 1934. He received his high school education at the Gymnasium Haganum in The Hague and obtained his Masters degree in Mathematics at the University of Leiden in 1959. After serving in the army for almost two years, he obtained his Ph.D. at the University of Amsterdam in 1964, with Jan Hemelrijk as advisor. In 1965, he was appointed Associate Professor of Statistics at the University of Leiden and promoted to Full Professor in 1968. He remained in Leiden until his retirement in 1999, while also serving as Associate Professor at the University of Oregon (1965), William Newman Professor at the University of North Carolina at Chapel Hill (1990–1996), frequent visitor and Miller Professor (1997) at the University of California at Berkeley, director of the Thomas Stieltjes Institute of Mathematics in the Netherlands (1992–1999), and founding director of the European research institute EURANDOM (1997–2000). At Leiden, he was Dean of the School of Mathematics and Natural Sciences (1982–1984). He served as chair of the scientific council and member of the board of the Mathematics Centre at Amsterdam (1983–1996) and the Leiden University Fund (1993–2005).

Bill served on numerous committees of the Institute of Mathematical Statistics (IMS), the Bernoulli Society for Mathematical Statistics and Probability (BS), the International Statistical Institute (ISI) and the American Statistical Association (ASA). For IMS, he was Associate Editor (1972–1980) and Editor (1986–1988) of *The Annals of Statistics*, and President (1991–1992). For the Bernoulli Society, he was President (1987–1989) and Editor-in-Chief of *Bernoulli* (2000–2003). He served ISI as Chair of the Organizing Committee of the Centenary Session at Amsterdam (1985), two-term Vice-President (1985–1989), program chair for the Session at Florence (1993), and President (1997–1999). He was a member of the Board of Directors of ASA (1993–1995). He was a member of the Corporation and the Board of NISS (1993–2002). He served as member and chair of the European Regional Committee (1969–1980) that organized the European Meetings of Statisticians, and for many years as an organizer of the Dutch Stochastics Meetings at Lunteren (1972–1999).

He gave the Wald Memorial Lectures (1992), the Hotelling Lectures at the University of North Carolina (1988), and the Bahadur Lectures at the University of Chicago (2005). He is an Honorary Doctor of Charles University at Prague (1997), a member of the Royal Netherlands Academy of Sciences (1979), and the Academia Europaea (1990). He is a recipient of the Van Dantzig Medal of the Netherlands Society for Statistics and Operations Research (1970), the Bernoulli Medal (Tashkent, 1986), the Peace Medal of Charles University (1988), the AKZO-Nobel Award (1996), and the Alexander von Humboldt Research Prize (2006). In 1996, he was made a Knight in the Order of the Netherlands Lion by Queen Beatrix of the Netherlands.

Bill has published some 80 research papers and two books.

*Rudy Beran is Distinguished Professor of Statistics, University of California, Davis, One Shields Avenue, Davis, California 95616-8705, USA e-mail: beran@wald.ucdavis.edu. Nick Fisher is Visiting Professor of Statistics, University of Sydney, NSW 2006, Australia e-mail: nickf@maths.usyd.edu.au.*

Our interview commenced on August 23, 2006, during the *Prague Stochastics* mid-week excursion, and continued over the next two days in Prague's old city, in striking rooms that once belonged to the nobility before Charles University acquired them.

## 1. CONTINENTAL EUROPE, ESPECIALLY THE CZECH REPUBLIC

**Interviewer:** Bill, since we're having this conversation during *Prague Stochastics 2006,* why don't we start with the Czech Republic and how you came to like it here so much.

**WRvZ:** I guess the first person I knew here was Hájek, whom I met a number of times, first at the 1965 Berkeley Symposium and later in Oberwolfach and Prague. The most memorable occasion was when he arrived in Oberwolfach by car with a trunkful of Czech beer. As the three of us all know, this is the best beer in the world, so this increased his popularity immeasurably, of course. But apart from a taste in beer, we had many other interests in common. One thing I remember about Hájek was a conversation that we had sitting around a table in Oberwolfach with Peter Huber, Hájek, and me. At one point, Hájek said, "You guys are real mathematicians; I'm just a simple insurance mathematician and I don't understand all these complicated things that you're writing about. However, I have a feeling that too many people right now are busy proving easy theorems in difficult spaces, and I think more people should prove difficult theorems in easy spaces." I've been quoting this to my students ever since.

I invited Hájek to come to Holland, but by that time, in the early seventies, he was already suffering from the kidney problems that killed him a couple of years later. I remember hearing him speak at Oberwolfach about his big theorem, the convolution theorem. I actually made notes, which is something I never do. You had the impression that this was a historic moment. He was crystal clear and gave you the feeling that you finally understood something that should have been obvious all along. His book with Šidák on rank tests was also fantastic: a real eye opener (Hájek and Šidák, 1967). And we should remember that he acted as the great interpreter of Lucien Le Cam's ideas in the early days.

**Interviewer:** So, Hájek was your first connection to Czech statistics? What happened next? Why did you keep coming back to Prague?

**WRvZ:** Our Czech friends were having a very difficult time and going there was almost the only way to see them. Also, I love Prague. It is the most pleasant and beautiful city in Europe. Over the years, I got to know quite a few people like Jana Jurečková, Marie Hušková, Zuzana Prášková, Petr Mandl, Václav Dupač, Jitka Dupačová, Jiří Anděl, Viktor Beneš, Jaromír Antoch, and many others. And ever since the velvet revolution, I feel happy and at home here. I also like the Czech self-deprecating sense of humor. A people that have the good soldier Švejk as their national hero, cannot possibly be bad!

**Interviewer:** Were you at the first Prague Symposium on Asymptotic Statistics in 1973?

**WRvZ:** No, I missed the first one. I went to all of the subsequent Prague asymptotics symposia. Interestingly enough, numbers two and three were not held in Prague, because for some reason the authorities wouldn't allow that and the organizers had some problems with the state travel agency Čedok running the show. In the provinces, there was less official interference and things were easier. So, the second Prague symposium was held in 1978 in Hradec Králové and the third in 1983 in Kutná Hora, the site of the old silver mines.

We actually went into the mine complete with helmets and such. The passages were very narrow, the ceiling was very low and the rocks were really rocky. We walked in a single file and Volker Mammitzsch was just ahead of me. He is a little heavier than I am and I am a bit taller, so we both got stuck repeatedly. Without the helmet, I wouldn't have had any brain left.

Another highlight of the Kutná Hora meeting was the first performance of the Kutná Hora choir. One evening we heard loud singing somewhere in the hotel where we were all staying. A quick investigation revealed that Mammitzsch and a number of his compatriots were singing German songs. We joined in the effort and as the group kept growing, it became more and more international. In the end, we were

---

> At one point Hájek said, "... too many people right now are busy proving easy theorems in difficult spaces, and I think more people should prove difficult theorems in easy spaces."

singing, "Yellow submarine" in Russian, which is not easy. I'm not claiming that our performance had great artistic merit, but it sure made a lot of noise. Finally, the hotel management appeared and asked us to stop and allow the neighbors some sleep. During the next few years, the Kutná Hora choir gave a number of memorable performances.

The fourth symposium in 1988 was held in Prague again and that's when they handed out medals to the faithful among their foreign friends and supporters. Rafail Hasminskii and I were honored with the Peace Medal of Charles University during an impressive ceremony in the Aula of the university. The end of the communist regime was in sight and with Rafa and me there was a nice balance between East and West. Ten other medals were awarded, two of which went to my former students and colleagues Kobus Oosterhoff and Roelof Helmers.

**Interviewer:** Were these meetings unusual for Eastern Europe?

**WRvZ:** Yes, the Information Theory Meetings and the Asymptotic Statistics Symposia were about the first locally organized conferences in Eastern Europe that attracted participants from the West. We got to know each other and built up mutual trust. Later, the Vilnius meetings in Lithuania had a similar effect for the Soviet Union. A large scale breakthrough occurred when the European Meetings of Statisticians which originated in the West, came to be held with some regularity in Eastern Europe. The first of these was in Budapest in 1972 and the second in Prague in 1974. But for me, Prague was where it all started.

**Interviewer:** How did your longstanding professional interactions with Marie Hušková start?

**WRvZ:** Marie was in Amsterdam very briefly in 1971, but in 1974, she first came to the Netherlands for a longer stay at Leiden and at the Mathematics Centre in Amsterdam on a Czech grant. The grant was pretty minimal, so at the Math Centre, we found some extra money for her to survive. We quickly became fast friends. She was definitely not a favorite of the communist regime. I believe that her problem was that she did not take a very active—or sufficiently enthusiastic—part in political matters at the university. She and her husband, Mirek, built their own apartment building together with a group of friends. This was allowed, but I figured it was probably frowned upon by the authorities as a capitalist idea. However, Marie tells me it was more like hard labor, working on the house on weekends and during the summer. Anyhow, like many other good scientists, her career did not advance very fast during the communist regime. After the velvet revolution in 1989, she immediately became vice-chair of the department and a full professor a few years later. She has been back in the Netherlands quite often and she is always welcome.

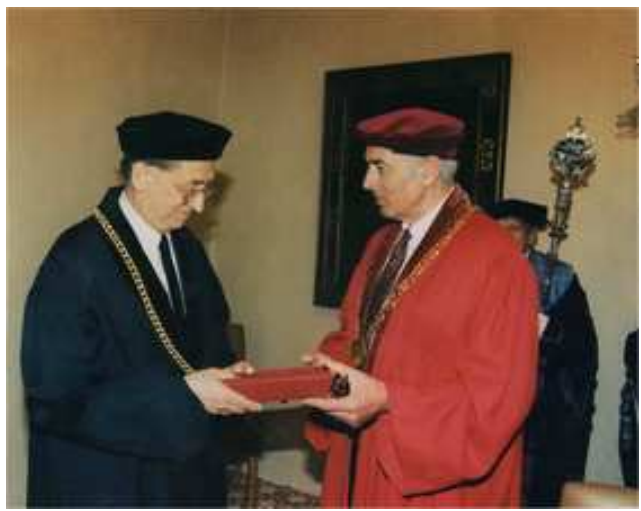

FIG. 1. *Vice rector Wilhelm hands over the replica of the gold medal after the honorary degree ceremony in 1997 in Prague.*

**Interviewer:** When did you become an Honorary Doctor of Charles University?

**WRvZ:** That was in 1997, long after the velvet revolution. The year before, I had a feeling that something was going on. Marie Hušková told me that it would be a good idea for me to meet Vice-Dean Netuka, apparently just to get acquainted. And then a letter from the Rector of Charles University arrived telling me that I was going to receive an honorary doctorate. That really bowled me over. It is something quite different from the usual fellow-of-this-that-and-the-other, and societies like IMS or ASA handing out fellowships are not quite in the same league as Charles University, founded in 1348. What is so nice about things like this is not the award itself, but the fact that your friends have taken a good deal of trouble to arrange this. I know they must have, because I've done this job myself at Leiden for Erich Lehmann's honorary degree. If I'm allowed one commercial, I recommend reading Erich's humorous account of this event in his new book *The Company I Kept* (Lehmann, 2008).

**Interviewer:** What was the ceremony like?

**WRvZ:** With the Rector's letter, there was a request for my measurements. I was in Berkeley at the time and Nancy Bickel was kind enough to measure me in various directions. Then a day before the ceremony, the Rector's gracious assistant Ms. Binova shows up at our hotel with an academic gown for me to wear. You are allowed to keep it and it has come in very handy. It is just as plain black as a Leiden gown, but made of thinner material, and hence easier to transport. I wear it in the Netherlands for Ph.D. exams in summertime or when I have to attend a ceremony at another Dutch university. Nobody has ever noticed the difference.

The ceremony itself was in the Aula of the university. It was really impressive. After 650 years' practice, they know exactly how to put on a great show. I ended by promising in Latin to uphold the dignity of Charles University, whereupon they put a chain with a gold medal around my neck. I gave a brief speech and then onward to a nice reception, where someone from the Netherlands Embassy succeeded in congratulating me twice! I think the person most impressed by all this, was probably my Italian graduate student, Marta Fiocco, a good Catholic who told me that never before had she been seated next to a cardinal (Cardinal Vlk of Prague in this case).

Before the ceremony, I was told about the gold medal and also that you were supposed to return it after the ceremony in exchange for a replica. They used this expensive piece of jewelry for every honorary doctor, and wouldn't want to lose it. Later, I learned that during the communist days, they handed out honorary doctorates to foreign Heads of State and assorted politicians. Haille Selassie was one recipient who apparently liked the thing and tried to walk off with it. I forget whether he actually succeeded or not, but since then they really watched you like hawks!

**Interviewer:** Any other interesting characters among the honorary doctors at Prague?

**WRvZ:** I'm not sure because I can't find a complete list. Simon Wiesenthal is certainly one, and also some quite impressive mathematicians like Kuratowski, Sobolev, Erdös, Atiyah, and Choquet.

**Interviewer:** When did you first visit the Soviet Union?

**WRvZ:** After I first met Albert Shiryaev in 1975 and started some joint work with Dimitri Chibisov around 1979, I got to know a growing number of colleagues in the Soviet Union. By the usual process of warnings against the "bad guys" received from the "good guys" I sorted out whom I would trust. I visited Moscow and Leningrad a number of times. That was a whole new experience. The first night I spent at the academy hotel in Moscow, there was a knock on the connecting door to the next room. When I opened the door, there was a Hungarian who said, "I have a bottle of palinka and I'm leaving tomorrow. Would you mind sharing this with me?" Much later that night, he said, "I have another present for you" and handed me a box of Kleenex. He added, "You won't know why I'm giving you this, but you will tomorrow." The Kleenex was fine, but then the toilet refused to flush. With some difficulty I acquired a piece of copper wire and fixed it. This made one Russian friend remark later, "Why are you people in the West so scared of Russian rockets? You know the toilets at the academy hotel don't work, so why should the rockets?"

At the Steklov Institute, I was graciously received by Yuri Prohorov. One evening, we had a superb dinner at Hotel Praha. The next morning I got up early with a somewhat wooly head to catch a plane back to Amsterdam. When I came down to the lobby at 7 a.m., there was Prohorov to say good bye.

At the Steklov, I also met Stacek Khmaladze, a Georgian who commuted between Tbilisi and Moscow, which is a distance of about a thousand miles. One day, he and I are walking in the streets of Moscow, on our way from A to B and pass a parked black car with the driver behind the wheel. Stacek stops, talks to the driver, and tells me, "Get in," and we are driven from A to B. I'm filled with admiration and ask how he does this. "Easy" he says and explains that (1) the car is black and, therefore, belongs to an important party member or official; (2) it is parked in front of an expensive restaurant and it is 1 p.m.; (3) it follows that the owner of the car is having lunch, which will take until 3 p.m.; (4) of course, the driver can use an extra buck.

My host at Leningrad was Ildar Ibragimov, who became a good friend over the years. To my pleasant surprise, I met Kagan in Ildar's office in Leningrad. Skorohod was also around and warned me to watch my tongue in the presence of a certain less desirable colleague.

**Interviewer:** Did you attend any of the meetings at Vilnius in Lithuania?

**WRvZ:** I attended two of the Vilnius meetings in 1985 and 1989. In 1985, I was interviewed by the Lithuanian radio station. They wanted to hear that everything in Lithuania was better than in Russia. It turned out that my necktie had the colors of the Lithuanian flag, so that really helped. In 1989, I believe that both Skorohod and I gave the opening lecture, which must have been the result of some delicate balancing act.

During the Vilnius meetings, the organizer Statulevicius used to throw a magnificent party at his dacha. At one of these occasions, when the party was well under way and the vodka was flowing liberally, two guys in black suits turned up and told me I was going to a dinner party at dean Bikelis' house. I told them, "No I'm not," and appealed to Statulevicius, but he told me there was no way out. So, they put me in a car and by the time we were back in Vilnius the vodka was taking effect. I arrived in great spirits at a very formal dinner party and I'm not completely sure my behavior was sufficiently decorous.

There was another memorable instance during the 1989 Vilnius meeting. At that time, the Soviet Union was about to fall apart and Statulevicius was heading a delegation that discussed the future status of Lithuania with Gorbachev. So, at a dinner party at Vygantas Paulauskas' house, we raised our glasses in a toast to Lithuanian independence. I remember that one person, who was generally considered to be the KGB representative at the Steklov institute, needed some persuasion, but he eventually joined us with a somewhat wooden face. Of course, Lithuania is now an independent state and a member of the European Union. A few years ago, Statulevicius has died, but on a recent visit to Vilnius a number of people told me they had heard him speak warmly of this international show of support of his colleagues.

**Interviewer:** What are your recollections of the Bernoulli Society World Meeting at Tashkent in 1986?

**WRvZ:** It was a great meeting in every respect. Also, the excursions to places like Samarkand and Buchara were lifetime experiences. One amazing thing that Albert Shiryaev pulled off was to find a nice compound with a number of bungalows set in a garden in the middle of Tashkent. It was used for conferences of local political VIP's. Apparently, nothing was going on there at the time of the meeting and Albert succeeded in getting hold of this place for housing some Bernoulli Society friends and speakers. Staying there also delayed the stomach problems that everybody was bound to have sooner or later. The food was just very risky.

Wherever we went, there was a Red Cross ambulance behind the buses. So, every once in a while, we would stop and somebody who was feeling bad would continue the trip in the ambulance. I believe one person actually spent a few days in a hospital. David Cox behaved heroically. During his talk, he suddenly left, but came back and finished his talk.

But the biggest test of the stamina of the Bernoulli Society faithful was a 45-minute speech by the Secretary General of the Communist Party of Uzbekistan about the blessings that communism had brought to the Uzbeks. The speech was in Russian (or Uzbek?) but when it was finished and the rush to the bathrooms was about to start, David Kendall produced a full translation into English that lasted another 45 minutes.

I think it wasn't generally known in the West at the time that Tashkent had been recently rebuilt, after being almost totally destroyed by a long series of earthquakes. One of the last days of the meeting we were sitting in one of the huge lecture halls listening to Paul Switzer speak about earthquakes. Suddenly I felt my chair moving a little and I noticed the huge chandelier on the ceiling swinging a bit. I looked around and saw other people do the same, obviously thinking, "Should we run, or not?" Luckily the movement stopped and we all stayed where we were.

The flight back from Tashkent to Moscow was also interesting. We boarded the biggest plane I have ever seen. It had two stories over its entire length, not just the front of it. I have never seen anything like that before or after.

## 2. EARLY DAYS

**Interviewer:** We know that you were born with a deep voice but we don't know whether or not you were muttering asymptotic expansions at the time. What was it that attracted you to statistics in the first place?

**WRvZ:** Well, it was mostly dissatisfaction with other things. After starting World War II by being bombed at Rotterdam, and ending it under artillery fire at Arnhem, I got into high school in 1945 and finished in 1951. I went to Leiden as a student. Everybody studied physics in those years. I guess we all wanted to know more about nuclear physics, in particular.

The problem was that we were taught by experimental physicists and their mathematics was a bit shaky. You had to major in two fields for the first three years, which in practice meant physics and mathematics. Then you could switch and major in either one for the remaining three years to your Masters degree. Our main math professor, Kloosterman, tried to instill some sense of rigor into us, and then the next lecture was by some physicist who made a mathematical mess of things. Even as freshmen we could see that. Then you had to do thirty-eight physics experiments and that really took a huge amount of time. Luckily, I had a good friend, my later probability colleague, Jaap Fabius, who has two right hands to complement my two left ones.

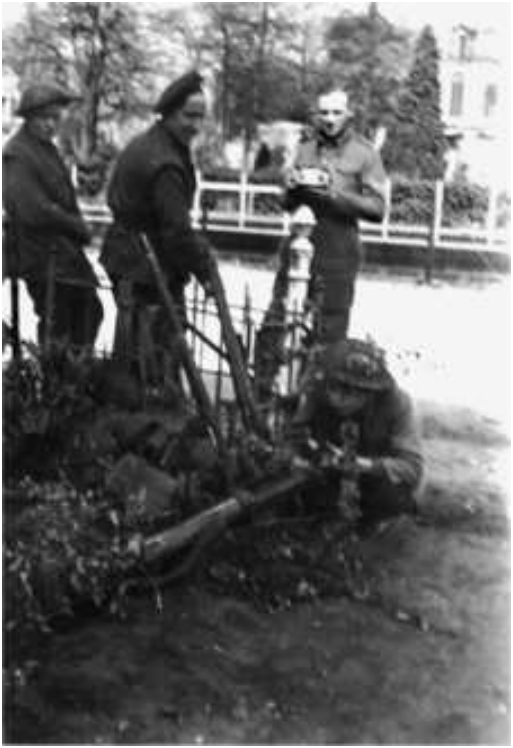

Fig. 2. *Bill at age 11 assisting the Allied troops at Arnhem in April 1945.*

**Interviewer:** So you were paired together?

**WRvZ:** Yes, I mostly I wrote down the results and tried to explain least squares to the physicists in our lab reports. This was the old Kamerlingh Onnes Laboratory, where they first liquefied helium way back when. For every experiment, we were required to produce two results that were sufficiently close, which is easy, of course. Unfortunately, by 1950, it was a pretty rickety place, so when a truck rolled by, the whole experiment had to be done again. To complete my misery, I made the mistake of taking chemistry as a minor, which meant another six months in the chemistry labs. After going through this for a year, I was so discouraged that I stopped doing anything at all and specialized in more pleasant student activities. Then, after three years or so, you were invited to tea by one of the physics professors to discuss your progress. I was told that my future was quite hopeless. My grades were okay, but there were not very many of them.

After about four years, I took hold of myself. I studied like crazy for a while and took my first degree after five and a half years instead of three. In those days, the university had no problems with that, and the main difficulty was to avoid being drafted into the army. The Dean of Students, who was supposed to help take care of such problems, was a good friend of mine. I chaired the student sports committee and he attended our meetings on behalf of the board of the university. He kept me out of the military, which was truly a miracle. By that time, I was really through with physics and switched to mathematics. In Leiden, this meant the pure variety. That was all very nice, but in those days majoring in math meant that you became a high school teacher and that didn't attract me very much.

Luckily, I found out that there was something called "Statistics" and that a correspondence course for industry was being taught by a statistical consulting firm in Rotterdam. I knew the people involved, so Jaap Fabius and I signed up to correct the written homework of the students of this course. We knew absolutely nothing and learned at the same time as our students did. We computed sums of squares, divided them by degrees of freedom, and called their ratio $F$ and pronounced the magic word "significant." It all sounded more or less reasonable, and the applications from industry were real and interesting. After a while, we got the idea that there might be some mathematics behind it all. The people involved in the course didn't provide much information about this, but all agreed that as mathematicians we would have a brilliant future in statistics without becoming high school teachers. So, we thought we had better investigate whether there would be any university teaching statistics. Obviously, Leiden was not doing this. We discovered there was someone in Amsterdam who taught something called the theory of collective phenomena, and further research revealed that this was probability theory. We went to this Professor Van Dantzig and he said fine, yes we could follow his courses.

**Interviewer:** What degrees did you have by now?

**WRvZ:** We had just got our first degree (called Candidate) which is halfway through the six years'

program for a Masters degree. For almost everybody, a Masters degree was the end of your university education. Nobody took a Ph.D. unless you wanted a career in research. A Masters degree was supposed to equip you for any other job after six years of study. I have to admit I needed $7\frac{1}{2}$.

**Interviewer:** So, the Ph.D. prepared people for an academic career?

**WRvZ:** Yes. That, and for work in the chic industries like Philips Electronics and Shell with prestigious laboratories.

**Interviewer:** How were you supporting yourself for all these years?

**WRvZ:** Up to your Master's degree you were supposed to be supported by your parents and make a little money of your own. In my last year, I had a job as an assistant in the biology department that paid relatively well. If your parents couldn't help and you had really superb grades in high school, there was a small number of scholarships.

**Interviewer:** Were the fees reasonable?

**WRvZ:** The fees were zero and we hardly used books, just lecture notes. So, it was just living expenses and we lived in really crummy places. You rented a room with a washstand and a coal stove from a landlady. I think there were only five thousand students in Leiden and many of them just commuted by train, living with their parents. It was considered a luxury to live in a crummy room in Leiden. But we had a lot of a fun. At the student's club, a beer was $9c$ US and at night they closed when the last guy left. Halfway through my studies I was beginning to doubt that I would ever finish.

**Interviewer:** When did you decide you wanted to do a Ph.D.?

**WRvZ:** Oh, I don't know. Jaap Fabius and I had this idea that we would go to Berkeley to get a Ph.D. That sounded great, but the guy who had kept me out of the army so far had disappeared in the meantime. So, the army grabbed me. The day after I got my Master's, there was a letter in the mailbox saying that I would report two months later at some army camp. It wasn't so bad. I mean it was a wonderful summer in 1959, and infantry training greatly improved my physique, which was in pretty dismal shape. Of course, I never got a Ph.D. in Berkeley. Jaap Fabius did.

**Interviewer:** You also spent time at a Navy lab?

**WRvZ:** Before the army grabbed me, I had heard a rumor that, while you were serving in the army, you could get a job at one of the defense laboratories. This was for physicists and chemists. Nevertheless, I applied to the appropriate government agency. I wrote what a great maths student I was and how much I knew about physics, and that it would be very good for them if they'd give me a job. For these laboratories, it was okay anyway, because they didn't have to pay these guys who were in the army. Of course, I got a letter back saying, who did I think I was, and no way were they going to do anything for me. That night, while having dinner with my future wife's parents, I tell them that I got this letter from this guy. My future father-in-law wants to know this guy's name and when I tell him, it turns out it is an old friend of his. So, he picks up the phone and says, "Pieter, my future son-in-law needs a job in one of your stupid laboratories and what has gotten into you not to give him one." So, there was another letter saying it was all a mistake and I'd get a job at the Navy physics lab after some months in the army. After six months in the infantry, the captain came to me one evening and said, "I have some papers for you. It seems you are going to The Hague to some laboratory. As an infantryman, you're not going to like this, and I can try to get you out of this if you want me to." I told him not to bother and happily arrived in The Hague. The first thing they told me at the Navy lab was that they didn't like to see army uniforms, so would I please come in civilian clothes. So, my army career effectively came to an end after six months, though I spent fifteen more at this lab. We did some interesting things there. Again, I learned some physics and I taught them least squares in return.

**Interviewer:** What was your first project?

**WRvZ:** After I arrived, they told me in no uncertain terms that they were unhappy with me because they had asked for a physicist and got a mathematician. So, I had better keep very quiet. Some genius who had been there before me and was now living it up at a NATO lab on a beach in Italy had developed a formula for the magnetic field of a mine-sweeping device. Because I was a mathematician, I would certainly love to calculate this field at various depths below the surface and for varying depths of the sea, with the aid of the most modern electric calculator. They handed me the formula and said compute it at twenty-five different depths over such and such an area. So, I pounded this ancient calculator for a couple of days and before I became totally unconscious, I discovered something strange. So, I said to my boss, "There is something weird; when you approach the bottom of the sea, the magnetic field goes to plus infinity. This doesn't sound right." He said, "That is none of your business, you just pound your calculator." Finally, I made some tables and graphs and somebody looked at them and started to get worried about this field going to plus infinity at the bottom of the sea. I asked to see the report where this wonderful formula was derived, but no, no, this was secret and there was no way I could see it. However, when the field was still going to infinity a few weeks later, they upgraded my security clearance and I finally got to see the report. It was written by an engineer from Delft and contained a glaring mathematical error. So, that established my credentials and I was allowed to do more interesting things.

**Interviewer:** Did you end up correcting that formula?

**WRvZ:** Sure enough. The error was actually easy to spot and to repair, because our measure theory professor Zaanen had been teaching in Delft before he came to Leiden, and always warned us to watch out for this particular mistake which, he said, engineers are liable to make. He was right and was delighted when I told him about my experience. After that, more interesting work turned up. I think the best achievement of our group was the design of a very efficient asdic set, which is equipment used to locate submarines by recording reflected noise. When my 15 months were up, I briefly considered a permanent job at the lab, but decided I wanted to get a Ph.D. In the meantime, I had married Lucie, Van Dantzig had died, and the student club in Leiden had burned down. I felt it was time to start a new life.

**Interviewer:** When you decided that you were going to study for a Ph.D., had you decided which university?

**WRvZ:** Well, I thought that at age 27 it would be a little late to go through the formalities of getting a degree in the US. In the Netherlands, statistics existed only at Amsterdam, but Van Dantzig had died. His most prominent former student in statistics was Jan Hemelrijk, who was a professor at Delft, but was taking charge of Van Dantzig's statistics department of the Mathematics Centre in Amsterdam. However, when you get a Ph.D. at Delft, you become a Doctor of Engineering and, after my aborted career as a physicist, that was the last thing I wanted. So, I went to see Hemelrijk and was relieved to hear that he would succeed Van Dantzig at Amsterdam. We agreed then and there that I would get an appointment in his Statistics Department at the Mathematics Centre (which is not part of the University of Amsterdam) and that he would be my thesis advisor. Hemelrijk got me the maximal possible starting salary, which my friends working for Shell or Philips considered to be a joke.

**Interviewer:** Tell us a little about Van Dantzig and Hemelrijk.

**WRvZ:** David van Dantzig was a topologist turned statistician during World War II, while he was hiding from the German occupation as a Jew. He believed statistics and applied mathematics would contribute in important ways to the development of society after the war. In 1946, he was one of the four founding fathers of the Mathematics Centre and head of its Statistics Department. Because he was basically self-taught in statistics and probability, there were obvious gaps in his knowledge. On the other hand, his early history in pure mathematics made him see and understand things that others would have missed. He taught me many things that few statisticians know about, but he didn't teach me many things that everybody else seems to know. Like other mathematicians in Amsterdam at the time, he had great interest in foundations. To my mind, his two most interesting statistical papers are Statistical Priesthood I and II, which are demolition jobs on Leonard Savage on subjective probability and Sir Ronald Fisher on fiducial inference (van Dantzig, 1957a and 1957b). Though he recognized the importance of applied statistics and laid down rules for correct statistical behavior that people at the Mathematics Centre were supposed to follow, he had no talent for applied work himself and wisely left this to Hemelrijk, who did have this talent and knew all the tricks. Hemelrijk had relatively little interest in mathematical theory, but greatly enjoyed a good applied problem and would delight in a novel and nonstandard solution. He and I got along famously

---

4AN EVENING SPENT WITH BILL VAN ZWET 9
from day one. In my first six months at the Centre, we spent most of our time together working on consulting problems. After these six months, he decided I could stand on my own feet, appointed me *sous-chef* of the department and doubled my salary. In practice, this meant that as a graduate student, I was basically running the department, relieving him of a lot of work. However, we still kept working together when an interesting applied problem turned up.

**Interviewer:** What kind of data were turning up in your consulting?

**WRvZ:** We had permanent contracts with a number of companies like Philips, Shell, the insurance business, and also some industrial consultant firms. They would show up any time and shoot questions at us. We knew these people well and such contacts were valuable for us. They also brought in some cash. Then there was an enormous amount of the usual scientific projects from all over the university, including a lot of Ph.D. thesis work. Usually, the Ph.D. student was sent to us by his advisor to find out what could be done with his data. In the early days of Van Dantzig, the answer was usually "nothing," because the wrong data had been obtained in the wrong manner and for the wrong reason. The Math Centre quickly became known as the "graveyard of medical science." Hemelrijk was more sympathetic and always willing to see what could be done with less than perfect data. These services were free of charge, unless we discovered that the research was financed by Bayer and supposed to establish the miraculous effects of Aspirin. Of course, there were the usual conflicts. If we told a Ph.D. student that there was really very little we could do with his lousy data, he would tell his advisor who would pick up the phone and call Hemelrijk. He would tell him that these kids at the Math Centre didn't understand the wonderful experiment he had designed for his student, but he felt sure that Hemelrijk would straighten them out. But Jan Hemelrijk was the perfect boss. He would tell his esteemed colleague that his "kids" were usually right, but that he would certainly look into this. His colleague would never hear from him again.

**Interviewer:** So, you began life as an applied statistician?

**WRvZ:** Yes, during the daytime anyway. At night, I would work on my thesis. During the day, I had a full schedule that also included some teaching.

**Interviewer:** Later I think you were best known as a theoretical statistician. When did the conversion happen?

**WRvZ:** I have always been interested in both theory and applications. However, consulting usually can't wait, so it gets priority and theoretical work can wait until you are home at night. As a result, I was literally working day and night, which is a bit much for a lazy person like me. Six months after my Ph.D., I went back to Leiden as an Associate Professor. When I got there, I felt I finally had some time to think without worrying about all of these chores.

**Interviewer:** What was happening with your nighttime work for your Ph.D.?

**WRvZ:** In the days of Van Dantzig, most people at the Centre used to write about rank tests. You would devise a test and prove asymptotic normality under the hypothesis. Of course, more imaginative people like Constance van Eeden went their own way and did something different. When I arrived, there was also a very nice and largely finished thesis on a test related to random graphs written by my predecessor as *sous-chef*, who had suddenly died. So, Hemelrijk decided that I might as well finish that piece of work. It turned out, however, that there were lots of loose ends so that this took much more time than I had thought (Bloemena, 1964). In effect, I think I wrote $1\frac{1}{2}$ theses instead of one.

In the meantime, I was, of course, talking to Hemelrijk about a topic for my own thesis. I didn't have any spectacular ideas and, as you can imagine, Hemelrijk wasn't terribly interested in the theoretical stuff that went into a thesis anyway. So, he had devised a system for being a Ph.D. advisor on automatic pilot, so to speak. Every visitor to the Centre of any scientific standing was asked to write a nice problem on a single sheet of paper, and these sheets went into a folder that he kept in his desk. So, I was handed this folder and invited to choose my favorite thesis topic. Most of the sheets didn't look very promising and the best I could find was a contribution of Gottfried Noether, who asked about an inequality for expected values of normal order statistics. It took me a few weeks to learn about order statistics and decide that this was a weird little problem. Then I learned about convex functions and proved the inequality. I went to Hemelrijk and told him I had solved this problem. He looked at it for a while and it was clear he was seeing the problem for the first time. He said, "Of all the unimportant things I have seen, this beats it." If anyone else had said that, I would have been mad, or discouraged, or both, but coming from Jan Hemelrijk, it didn't bother me much. I decided that from now on we would be friends, working together on applied problems, but I would do my theoretical research on my own.

This arrangement worked wonderfully well. I realized that if you wanted mathematical guidance, then Hemelrijk was not the person to go to. But I didn't and I loved being left alone. I generalized the set-up considerably and added lots of new applications and that really produced interesting results. In the end, my thesis (van Zwet, 1964) was totally unrelated to anything else going on at the time, but looking back, I still think it has considerable charm and I like it. Surprisingly, the Math Centre sold something like 800 copies and the topic resurfaces every ten years or so, with new results being added. In the end, I think that even Jan Hemelrijk liked it, though he never really read it. A few weeks before my Ph.D. exam he said, "I trust you are sure there are no serious errors. You see, I don't want any problems with my colleagues."

**Interviewer:** What was happening in Leiden in those days?

**WRvZ:** Nothing very exciting. In 1968, I was promoted to Full Professor because I'd had an offer from Eindhoven in 1966. You may think it strange that this took two years, but such appointments had to be signed by the Queen and mine was signed at her summer address in Italy. In 1968, Hemelrijk organized the European Meeting of Statisticians in Amsterdam. Of course, I had to do quite a bit of the work, but the fact that I was no longer in Amsterdam helped a little. Two weeks before the meeting started, the Russians invaded Czechoslovakia. We had long discussions on whether the meeting should be canceled, but decided this was just plain impossible.

In the spring 1969, Jan Hemelrijk told me one day that, having organized the European Meeting in Amsterdam, he felt he'd done his bit and resigned from the committee responsible for these meetings. He had told them that I would be his successor. After a while, it became apparent that I had inherited a committee that was dead and organized nonexistent European Meetings of Statisticians.

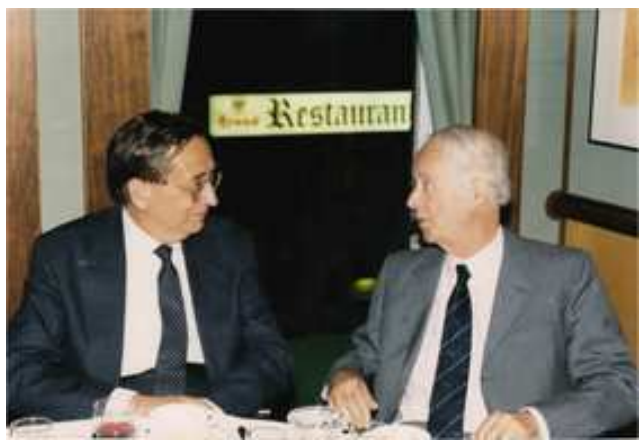

Fig. 3. *Dinner with Bill's thesis advisor Jan Hemelrijk at Amsterdam 32 years later in 1996.*

**Interviewer:** Is this what turned you into an activist?

**WRvZ:** Yes, this is something I realized only recently when I was clearing out my office after my retirement in 1999. When you're throwing away a lot of stuff and start reading old letters and documents, you see what you have been doing with your life, and I suddenly discovered that I had become very active in all sorts of things around 1970. Then I remembered that during my first years in Leiden, I was just doing my duty and getting a little bored. A Full Professor's salary in the Netherlands was fixed with only cost of living increases each year. You could do whatever you wanted and nothing would change. Nobody seemed interested in what you were doing—or whether you were doing anything at all—either. I felt I had reached the end of my career in 1968 at age 34. I decided that if I wouldn't start moving things myself, nothing would happen.

## 3. USA

**Interviewer:** How did your visits to USA start?

**WRvZ:** Fred Andrews spent a year at the Mathematics Centre while I was there. He was my first link to the US. He was very nice and talked to us a lot. Then after I got my Ph.D., he invited me to come to the University of Oregon in Eugene for six months. This caused some problems because my appointment in Leiden would start on January 1, 1965, they wanted me in Oregon on January 2, and I really

---

"I trust you are sure there are no serious errors [in your thesis]. You see, I don't want any problems with my colleagues."

> "Well, this university has done quite well for four hundred years without teaching statistics and I guess six more months won't hurt."

wanted to go. At a reception, I met the Secretary of Leiden University, who was a very nice fellow, and decided to take the bull by the horns: "What would you say if I asked for leave of absence the day I start here? I would like to go to the US for a semester." He looked at me and said, "Well, this university has done quite well for four hundred years without teaching statistics and I guess six more months won't hurt." My stay at Eugene was great. I started working on entirely new topics, learned a lot from Don Truax and made new friends.

So, the first time I set foot on American soil was in 1965. At Eugene, they made me teach two courses during the first quarter, both at eight a.m. That time of the day is not my best.

One was elementary statistics and the other one was on differential equations. This I really don't like and I just kept one step ahead of the students. What was new to me was girls walking into my office, telling me that they were afraid they would fail my course, and then start crying. When I asked Fred what to do about this, he said, "Nothing, except keep your office door open at all times."

In the elementary stat class, there was a group of really big guys who faithfully handed in their homework that was disastrous. Their exam papers were equally bad, so I flunked the lot of them and gave them an F. This earned me a visit from the friendly Chair of the Math Department, who explained that I had just flunked the Oregon football team. As they needed a certain number of grade points to be classified as bona fide students, the future of the U of O was in danger. I seemed that their coach had seen the word statistics somewhere and decided this course would be a good idea to help them keep the score during their matches. I agreed that this disaster should be averted and discovered some convincing reasons to raise their grades, provided they wouldn't come back for the next course. This made everybody very happy.

Then one day Fred Andrews said, "You have to go to the Fifth Berkeley Symposium," and I replied, "Yes I'd love to, but I have no money so what do I do about this?" So he picked up the phone and called Betty Scott. He said, "Betty, I have a deserving young man here who wants to come to the Symposium. Do you have any money left?" Betty said, "I have five hundred bucks left in a box somewhere and he can have it." My eyes were popping out. So, Lucie and I got into our car and drove off to Berkeley.

I remember that when I arrived the first person I met was Erich Lehmann, who of course not only said he had read my thesis, but actually gave some evidence of having seen it. The next day I met Herman Chernoff. We got talking and he said, "What are your interests?" I replied "convex functions," and he said, "That is an interesting hobby for a statistician." The great sensation during the Symposium was the arrival of the Russians. Someone was lecturing about something and suddenly the door of the lecture room opened and there was loud applause. The speaker looked confused, wondering whether he had said something sensational. But it was just Yuri Prohorov sticking his head into the room.

**Interviewer:** A little later, you spent an entire summer in Berkeley, didn't you?

**WRvZ:** Yes, I had met Dick Barlow. We noticed that the work on reliability theory that he and his friends at Boeing were doing was somewhat related to my thesis work for the special case of the exponential distribution. What is called a distribution with increasing failure rate in reliability theory corresponds to an ordering with respect to the exponential that I had studied. Dick was in the Operations Research Department in Berkeley and invited me in the summer of 1967. It was a pretty stimulating visit. We wrote a really nice paper together (Barlow and van Zwet, 1970) and drank a lot of stuff called Orzata.

My next visit was during the last Berkeley Symposium in 1970. To my total surprise, I received an invitation to give a talk at the Symposium, and they even bought airline tickets. I knew more or less what went on during the Symposium after attending in 1965, but giving a talk was definitely something else. The symposium was held in Stanford for one day, and this happened to be the day my talk was scheduled. Before my talk, someone pointed out that George Polya was sitting in the front row and that unnerved me a little. However, at the end of my talk, Polya was still sitting there and nobody else tore me apart either. So, I decided to chalk this down as a success.

I came back to Berkeley in 1972, again on Dick Barlow's grant. This time Lucie and our two boys accompanied me and we stayed for almost six months. Dick had found a beautiful house for us in Rockridge. I bought a bike and rode down College Avenue to the campus. This was really dangerous because Americans are not used to bikes and open the door of a parked car right in your face. Somehow I survived.

Just before I came, I got a letter from Kjell Doksum. He was involved in the IMS Regional Meeting in Seattle that summer, where Peter Bickel would give an invited talk on asymptotic expansions for distributions of rank statistics. This was a new topic and would I be interested to be a discussant? I wrote back, "Yes, I would certainly be interested because a Ph.D. student and I are obtaining such expansions at this very moment." Peter and I knew each other a little, so we got in touch. I think he found it hard to believe that we had actually done these things, and asked whether we meant that we could write down the expansions—which anyone can do— or prove their validity. I said we could do the latter for one-sample rank statistics and added that that included Wilcoxon's statistic. This he obviously found too much to believe, because according to folklore, you can't deal with lattice statistics and Wilcoxon is of that type. Anyhow, a few days after we arrived, I was explaining what we'd been doing in Erich Lehmann's seminar. I think I convinced people that this could be done, but we also agreed that this was just a first step in a very difficult and time-consuming program. Both Peter and I became interested in this topic because Hodges and Lehmann had pointed out in a paper why this was important (Hodges and Lehmann, 1970). So, Peter and I joined forces and worked like crazy for six months, with Erich looking over our shoulders, reminding us all the time that we should really do a complete job and not leave any loose ends dangling. It took us a few more years to finish, but in the end we published our results in the longest and the third longest papers in the Annals (Albers, Bickel and van Zwet, 1976 and Bickel and van Zwet, 1978). I'm afraid only the truly fanatic have read them. During the next 25 years, I visited Berkeley frequently to work with Peter.

**Interviewer:** How did you find the Statistics Department at Berkeley when you visited it?

**WRvZ:** Really great. You see, it is much better to be a visitor in a competitive American environment, than to be part of the regulars. As a visitor, everybody is nice to you. You are not going to be around forever and you are not competition. So, I honestly couldn't point at anyone who wasn't good company.

When I first came in 1965, I was a little scared of Betty, because I had the feeling that having an extra cookie in the coffee room might be the end of me. I got over that later.

Neyman was always the perfect European gentleman to me. Let me illustrate this with a little story. As I said a moment ago, there was this IMS Regional Meeting in Seattle at the end of our stay in Berkeley in 1972. After that we would fly back home. Anyhow, I am clearing out my desk and I go to say goodbye to Betty and thank her for her hospitality of having me in the Department. She says, "Oh Bill I didn't know you were leaving I thought you were coming back after the Seattle meeting," so I tell her no, we'll be going straight back to Europe. When I get home, the phone rings and it is Betty saying, "I'm sorry I didn't realize you were leaving, would you come and have dinner at Mr. Neyman's house tonight with your wife?" I say, "Betty I'll be delighted, but I'll have to call round to see if I can find a baby sitter, because we have two small boys." Betty says, "Oh no problem, you just bring them along." I say, "Betty you don't know what you are saying, they are one and three years old," but Betty says, "No problem. Peter Clifford is also coming with his wife and small children." So, what can you say? We pack the kids into the car and drive to Neyman's house. We have cocktails and Neyman starts speaking French to me, as he used to do.

The idea was that the grown-ups would have dinner in the dining room. In the living room, they had set up a low table with small chairs where these kids were supposed to eat. So, after we went into the dining room, the kids presumably started rubbing spinach into each other's hair; they were really a riot. At one point, one of the two mothers got up but Neyman said, "We do not get up during dinner," so the kids kept screaming and yelling and throwing spinach around. So, the moment the dinner was over we left and then one of our kids started complaining that he left one of his toy cars in Mr. Neyman's chair and wanted us to go back and fetch it. We said, heartlessly, "Forget it."

Neyman was always very nice to me, but I had the feeling it would not be a good idea to contradict him. Luckily, he wasn't always listening very carefully to what you said. One day when I walked into Evans Hall in Berkeley, I ran into him. He said he hadn't realized I was around and we chatted for a few minutes. After I got to my office, the phone rang and Neyman's secretary informed me that Mr. Neyman would appreciate my attending his seminar that afternoon. I realized that this was not a matter of choice, so I went up to the 10th floor to attend the seminar. I hid in the back of the room, but of course Neyman had spotted me and came over after the talk, saying, "Don't the Dutch get thirsty at 5 o'clock?" I could hardly deny that, so off we all went to the Faculty Club. After a few drinks and the Polish toast, "To all of the ladies present, and some of the ladies absent," I thought it was time to leave and noticed other people putting this idea into practice. However, Neyman had decided that Grace Yang and I were going to have dinner with Betty and him. Even though I probably had other plans for the day, I can only say he was a gracious host.

**Interviewer:** Anyone else at Berkeley you'd like to mention in particular?

**WRvZ:** I mentioned Erich Lehmann before. Through his books, he has had an enormous influence on statistics. If people now say that optimality is out, let me tell you it will be back in one form or other.

Wassily Hoeffding, Lucien Le Cam, and Erich are my three heroes, who most influenced my thinking. All of us have, at one time or another, been indebted to Lucien for the depth of his insights. I used to step into his office now and then with a question. He would always take time to reply in great detail and would be really enlightening.

For many years, I invited Lucien to come and speak at the annual Dutch statistics meeting at Lunteren but somehow this didn't happen. Then he couldn't fly because he had trouble with the pressure on his eardrums. Finally, this got solved by the same device they use for children with frequent ear infections. They put tiny tubes through your eardrums. That problem being solved, Lucien agreed to come to the 25th Lunteren meeting in 1996. At the meeting, he delivered the best talk I ever heard him give. For some reason, he only had the odd-numbered transparencies with him. Nobody knew where the even-numbered ones had gone, but apparently that didn't bother him at all.

I should say a few words about Peter Bickel. Peter is my best friend among my professional colleagues. Working together, we complement each other beautifully. Peter has read a lot, which is something I try to avoid. He writes very well, but has trouble multiplying two and three, whereas I know that the answer is seven, but check and re-check everything I do. We have collaborated on and off for 35 years now. Some people say that after all these years we are actually beginning to sound the same. During my retirement ceremony at Leiden, he gave a speech in the university auditorium and compared my work with that of Wassily Hoeffding, which is the biggest compliment I ever got. What else can I say? Peter is my friend. Apparently, Queen Beatrix of the Netherlands agrees with me, because she made him a Commander of the Order of Orange-Nassau. This is not something that happens to many people.

**Interviewer:** How did your relationship with the Statistics Department in Chapel Hill come about?

**WRvZ:** The first time I was in Chapel Hill was in 1979 when I attended the symposium in honor of Wassily Hoeffding on the occasion of his 65th birthday. My next visit was in 1989 when I gave the Hotelling lectures, and after that I visited on a regular basis between 1990 and 1996.

When I was there in 1989, they gave me Wassily's old office where I discovered a copy of his famous unpublished paper on what I had christened Hoeffding's decomposition (Hoeffding, 1961). In fact, I wrote about this paper in an introductory article in the volume of his collected works (Fisher and Sen, 1994), but later I discovered that the paper itself was not in this volume, because the editors figured that if Wassily hadn't published it, then it shouldn't go into his collected works either. So, in a little while, mine will probably be the only existing copy of that seminal paper!

When I was in Chapel Hill again in 1990, Ross Leadbetter, who used to visit Wassily regularly and took care of all sorts of things for him, took me along a couple of times. That was the last time I saw him. He died in February 1991.

I already said earlier that I'm one of Wassily's greatest admirers. We first met in the peace and quiet of Oberwolfach in the early 1970's. Talking to Wassily was a slow business at best, and you had to ask a direct question if you wanted an answer. This made our second meeting a little difficult. We were at a party during an IMS meeting, sitting right next to an extremely loud band. I wanted to ask him something but that was hopeless.

Wassily's papers are beautifully polished and just deal with the main issues of a problem area. He was the statistician's statistician, in the sense that he happily left it to others to investigate the consequences and applications of his results, which were many. He always kept things as simple as possible. Also, you rarely find, "There exists a constant C such that...". in his papers. No, he'll give you the number—one of the truly great men in statistical science.

## 4. CONTRIBUTIONS TO PROFESSIONAL SOCIETIES

**Interviewer:** Let's talk about the people who got statistics going in continental Europe.

**WRvZ:** When I was a student, I guess Harald Cramér's group in Stockholm (including Carl-Gustav Esseen) was the only one in continental Europe with international acclaim in statistics as well as probability. In the Soviet Union, there were excellent probabilists. But statistics was starting in Denmark, the Netherlands, France, Italy, Austria, Czechoslovakia, Hungary, and Rumania. In Volume I of the Fourth Berkeley Symposium in 1960, you find the names of Dalenius, Hájek, de Finetti, Dobrushin, Fortet, Rényi, Schmetterer, Špaček, Vincze and Wold among the authors. Admittedly, many of them would be classified as probabilists, but something was going on in their countries.

**Interviewer:** What books were being read?

**WRvZ:** As students we were using Cramér's *Mathematical Methods of Statistics* (Cramér, 1946). I guess that M. G. Kendall's *The Advanced Theory of Statistics* (Kendall, 1948) was also used a lot. Then Erich Lehmann's book *Testing Statistical Hypotheses* (Lehmann, 1959) appeared. That was a real eye-opener. Scheffé's *The Analysis of Variance* (Scheffé, 1959) also had a profound effect on me. I finally understood all of these sums of squares that had bothered me for so long. As general texts, Wilks' *Mathematical Statistics* (Wilks, 1962) and Rao's *Linear Statistical Inference and Its Applications* (Rao, 1965) were popular for a while. Of course, Schmetterer's *Mathematische Statistik* (Schmetterer, 1966) was really solid, but even reading the statement of some of the theorems was hard work. For general probability, Feller's *An Introduction to Probability Theory and Its Applications, Vol. I* (Feller, 1950) was—and still is—the most beautiful way to approach the subject. For a more abstract account, we used Loève's *Probability Theory* (Loève, 1955). When Feller's *Volume II* (Feller, 1966) appeared, I first learned of the existence of expansions that would later keep me busy for a full decade. For a general audience, there was Yule and Kendall's *An Introduction to the Theory of Statistics* (Yule and Kendall, 1950). My dad, who was a lawyer, owned a copy that is still somewhere on my bookshelves.

In Europe, we had almost no chance to meet any of these celebrities in person, because there were hardly any international statistics meetings. During my first year at the Math Centre, I went to the ISI Session in Paris in 1961. Lucie and I had been in Rome on vacation and went to Paris by train. In our compartment, there was a gentleman who told us he was also going to attend the ISI Session, so I thought we'd come across a kindred soul. He went on to explain that mathematics was a lot of nonsense. All you needed was to collect data and they would speak for themselves. This was my first acquaintance with official statistics and it came as a bit of a shock. Much later I found out that most government statisticians have a broader outlook.

One of the nice things about the ISI session in Paris was that I got to know Bart Lunenberg, director of the Permanent Office of ISI. As I got more and more involved in ISI, Bart and I would work together during the next quarter century. He ran ISI beautifully and he was a great communicator. He had the gift of making all important decisions himself while giving others the feeling that it was really their idea. He was always willing to spend time with young people, which was a lot of fun. About ten minutes after we first met at a reception that was slightly less lavish than most, Bart said to us, "I think things are winding down here. Let's go look somewhere else" and we crashed a party next door in the same building.

During the Session, there were one or two fierce discussions. The one that impressed me most was a lecture by Alan Birnbaum, who claimed he could reconcile the ideas of Fisher and Neyman. After he had finished his talk, Fisher got up and disagreed. The gist of his remarks was easy: Fisher was right, and Neyman was wrong. How could anyone in his right mind try to reconcile their ideas? While this went on, Neyman was sitting in the front row, looking at his shoes. I really admired him. Apart from that, my main discovery during the Session was that these celebrities whose papers I had read, really existed as real live persons that you could talk to. I actually first met Neyman and Betty Scott during an evening boat trip on the Seine, with lots of good food and wine. All in all, the parties were more memorable than the statistical theory at the Session itself. However, other international statistics meetings simply didn't exist in Europe.

**Interviewer:** But later there were the European Meetings of Statisticians. How did they get started?

**WRvZ:** That was indeed a great accomplishment. Jim Durbin had this idea that there should be an annual meeting in Europe on mathematical statistics and probability. Henri Theil, an econometrician from Rotterdam and later Chicago, had similar ideas for econometric theory. ISI wasn't interested, but Jim turned to IMS and proposed that they should organize regional meetings in Europe. They got strong support from IMS president Erich Lehmann, and in 1962 the first European Regional meeting of IMS took place in Dublin. This was truly a great moment for statistics in Europe. Because the meeting started very informally with a session of contributed papers, I happened to be the very first speaker at the very first European Meeting of Statisticians. It was my first talk at an international meeting and turned into a somewhat traumatic experience. Jan Hemelrijk had told us students that since nobody can read what you write on a blackboard in a large classroom, we should prepare a one-page handout for people to read while we were speaking. This is a really bad idea, because you lose your audience. I already suspected that that might happen, but things took a different turn. Hemelrijk operated under two side conditions. First, he would never enter an airplane, and second, he would never set foot in Germany. Given his history as a resistance hero during World War II, the second part was easy to understand; the first was more complicated. Of course, this made lots of places difficult to reach, and Dublin was one such place. He would go by car and, therefore, was going to transport large numbers of handouts for all of us. He would start out with his car on the North Sea ferry, drive across England and then onto another ship to cross the Irish Channel. The final part of the trip went wrong—they couldn't take his car, or something like that—so he arrived at the meeting five minutes in advance of my talk without the sacred handouts. For me, it was a valuable lesson in improvisation and I suspect I made a bit of a mess of it. Happily, Z. W. Birnbaum, one of the kindest people on earth, was kind enough to say something nice about my talk.

Everybody felt that the Dublin meeting had been a great success and the European meetings followed each other in quick succession: Copenhagen (1963), Bern (1964), London (1966), Amsterdam (1968), Hannover (1969). There, and at Oberwolfach, I got to know my contemporaries like Ole Barndorff Nielsen, Søren Johansen, Frank Hampel, Peter Huber, John Kingman, Klaus Krickeberg, Jef Teugels, Flemming Topsoe, and many others.

But something went wrong at Hannover. I said earlier that in the spring of 1969, Hemelrijk had put me on the European Regional Committee (ERC) of the IMS that organized these European Meetings. Unfortunately, I was unable to attend the Hannover meeting, so I sent a proxy to attend the meeting of the ERC. He came back with a rather alarming report. It seemed that the committee consisted of a senior statistician or probabilist from various European countries, and that after each European Meeting, they decided which one of them would organize the next one. This person then also became the committee's chair. So, presumably at Amsterdam, they had decided that Klaus Krickeberg would organize the next European Meeting at Hannover. Klaus had told them that this was impossible because he would be somewhere else—I believe South America—at the time, but apparently this wasn't felt to be a problem. So, nothing much had happened and the meeting mainly went through because it was joint with the Biometric Society, which was better organized. The local biostat professor called a meeting of the ERC, but few people attended and nothing was decided. So, it looked like the European Meetings of Statisticians had died.

**Interviewer:** You said earlier this turned you into an activist.

**WRvZ:** Yes, I felt strongly that the European Meetings of Statisticians were very important for the development of statistics in Europe and reviving them became my first project in this new mood. I started by writing to all members of the European Committee, introducing myself as Hemelrijk's successor, telling them what went on at Hannover and raising the question what we were going to do about this. I received hardly any reaction and concluded the thing was truly dead. Correspondence with Klaus Krickeberg made it clear that he wouldn't like to continue as chair. What to do next? My general rule of behavior was, "When in doubt, ask Ingram Olkin, the walking encyclopedia of the profession." Ingram told me, "If you need an infusion of energy in Europe, get Joe Gani." So I wrote to Joe, explaining the situation and asking him if he'd agree to chair the committee if I could get the other members behind this idea. Joe immediately agreed, so I wrote to the committee members proposing Joe as our next chair. I added that no reply would be counted as a "yes," which got me a critical letter from Peter Huber who agreed, but complained about my lack of democratic attitude. I have heard this remark more often, but it doesn't bother me.

True to Ingram's prediction, Joe started moving with tremendous energy. Most other members of the ERC dropped out of sight, so Joe and I had our hands full. We got an offer from Jean-René Barra to organize a European Meeting in Grenoble, but then something really surprising happened. Through Lunenberg's ISI office, we received a letter from a Hungarian Academician whom we knew by name, offering to organize a European Meeting in Budapest in 1972. So, here was suddenly the scientific, nonpolitical link to Eastern Europe, and a chance to meet a large number of our colleagues there: in 1972 statistics saw its first fully international meeting in Eastern Europe, organized by our colleagues and not by some political agency. Once the Hungarians came through, we had another European Meeting in Eastern Europe in Prague in 1974, and the Grenoble meeting took place in 1976. By that time, the ERC had been rejuvenated and the European Meetings were well-established scientific events again. Joe Gani had left, but I found Jim Durbin willing to return to his old love and chair the European Committee for a few years.

**Interviewer:** Was there general agreement to meet in Eastern Europe in the middle of the cold war?

**WRvZ:** Of course not. The standard way of thinking in those days was that you should not go to communist-run countries or collaborate with people there because the authorities in these countries would view this as a sign of recognition, and exploit it in some way. A similar tactic was to say you would come if the regime took certain measures that you were sure they wouldn't. Of course, the result was the same. However, having witnessed the destruction in places like Budapest in 1956 and Prague in 1968, and talking to refugees, it became difficult to believe that Eastern Europe was an area filled with supporters of communist regimes. Also, we began to meet fellow scientists from these countries. First they'd show up at a meeting shepherded by someone representing their regime, but after a little while it became easier to talk to each other. Once you use that opportunity, you find out amazingly quickly who are the good guys and who are the bad guys. The good guys will warn you against the bad guys, and the rest is just a matter of comparing information from different sources. Of course, the definition of "good" and "bad" is not perfectly clear, but after the collapse of the communist regimes, we found that we hadn't made a single major error in our judgment of the people we were dealing with. Some of them may have been only moderately okay, but we knew that, too.

The next question is, "Why should you worry about people in these countries at all?" The answer is that a scientist suffers from being isolated. Without a chance to travel and meet colleagues, with very little access to foreign literature and required acceptance of the prevailing political doctrine, life of a scientist is pretty dismal. If something like an international community of scientists exists, it should try to help their members in such circumstances.

So, there is the difficult decision of either risking supporting a malevolent regime, or not helping deserving colleagues. The rule we developed to make this decision is as follows. If the bona fide scientists (that is the good guys) ask you to visit their country for a meeting or otherwise, then you should go. This rule should, of course, also be applied to countries with dictatorial right-wing regimes. Of course, no rule is perfect, but I think this one has served us well, and the first people we started talking to in Eastern Europe and the Soviet Union are still our friends today.

---

> My general rule of behavior was, "When in doubt, ask Ingram Olkin, the walking encyclopedia of the profession."

There was another interesting thing with the communist regimes in Eastern Europe. They were never entirely predictable. I remember that I wanted to invite Boris Levit from Moscow to visit the Netherlands. This appeared impossible because he had applied for visa for Israel. But Boris suggested inviting him by writing directly to the KGB every six months or so. I would never get an answer, but perhaps something would happen. So, I found out that Kacha Dzaparidze's wife in Amsterdam had a Russian typewriter and she typed a convincing-looking Russian letter. We did this every six months for a number of years and nothing happened. Then one day, out of the blue, Boris suddenly showed up and wisely decided to stay.

Another, even more unexpected instance, was Pál Révész's officially approved move from Budapest to Vienna. One day, Pál discovered that the vestiges of the old Austro-Hungarian Empire still existed, in that some official rules from those days had never been retracted. He found out that it would perhaps be possible to accept a position in Vienna, the old imperial capital. There were only fairly mild rules that you had to follow. So, he first got himself a job offer from Vienna and then he went to the right department or ministry in Budapest and told them he was going. They probably looked somewhat flabbergasted, but when they looked up the law, they found there was very little they could do about it. Of course, this process took many months, so Pál thought it would be best not to be around to answer awkward questions. He spent six months with me at Leiden until the dust had settled and he was permitted to go. As a Hungarian citizen officially living in Vienna, he had the best of both worlds. He was beyond reach of the Hungarian communist regime, but getting his car repaired was a lot cheaper in Budapest.

**Interviewer:** Let's go back to your new activism around 1970. Do the meetings at Lunteren come under that heading?

**WRvZ:** Like my involvement in the European Meetings, unhappiness with the state of affairs was also the cause of this second burst of energy. In 1971, Ron Pyke was spending a semester in London and I invited him to come to the Netherlands for a week. The standard sources of money for such a visit were some funds of the Ministry of Education administered by the Mathematics Centre. They were happy to finance this trip, but there was a rule that said that you had to let your colleagues at other universities know about this visit and ask them if they wanted the visitor to give a talk in their department, too. I dutifully did that and all of my colleagues said yes, they would like to have this great man come over and give a talk at their place. So, Ron and I spent most of the week in my car and in hotels all over the country, to hear Ron speak at five different places to a hastily assembled audience of 3 or 4. I thought this was plain idiocy, and felt it would be more efficient to assemble the audience in one place for a few days, and invite something like six speakers people wanted to hear during the year at this place and time. It would have an additional social function of annually giving the math stat and prob community in the Netherlands a chance to meet. I, for one, had no idea who they were. Being the most junior stat professor in the country by a large margin, I first got the support of Theo Runnenburg, the slightly senior professor of probability at Amsterdam. Then I wrote to my other colleagues proposing this plan and they all replied, "Of course this is not going to work, but feel free to try, young man." So, we held the first meeting at a conference center in the woods in a place called Lunteren in 1972.

**Interviewer:** How did this first meeting turn out?

**WRvZ:** Well, first of all, the meeting very nearly didn't happen. On Sunday night, before the meeting started on Monday, we had the biggest storm in the Netherlands in decades. We were staying in a small hotel nearby, together with the speakers. The top floor of this hotel was a wooden structure that had obviously been added on later. Brad Efron and I had our rooms on this floor and during the night the entire floor was swaying back and forth. At one point, we both decided to get out of our rooms and met in the corridor. There wasn't much we could do, so after a while we went back to bed. The next morning 750,000 trees had come down in the area where we were and the roads were blocked in many places. Somehow we succeeded getting to the conference center and so did the participants. The meeting actually started right on time!

We had a great group of speakers: Hermann Dinges, Brad Efron, Peter Huber, Søren Johansen, Joop Kemperman, and Jef Teugels. Each of the six speakers gave two one-hour talks and there was a lot of discussion, some of it pretty fierce. In particular, Hermann Dinges can always be relied on to produce some fireworks. During Brad Efron's talk about the James—Stein estimator and shrinking,

Hermann got up and said, "You don't really believe that yourself do you?" This kind of attack is unusual in the US and Brad wasn't quite sure how to handle this. He did all right, though, and Hermann is a nice guy who just wants to start a serious discussion. In general, the meeting was pretty interesting and the idea of keeping people together in he middle of the woods for two and a half days with the bar open until 1 a.m., gave a lot of opportunity to interact. I was amazed by the large number of about sixty participants. We took a vote at the end of the meeting and it was unanimously agreed that Theo and I should organize another meeting the next year. In fact, the annual Lunteren meetings became a fixture and, in 2007, we had the 36th installment. I quit as an organizer when I retired in 1999, and now Mike Keane and Richard Gill are responsible for these meetings. Throughout these 36 years, hardly anyone turned down our invitation to speak at Lunteren and the list of past speakers reads like a *Who's Who* in probability and mathematical statistics.

**Interviewer:** Turning to some of your activities with professional societies, you've been President of the International Statistical Institute, the Institute of Mathematical Statistics, and the Bernoulli Society. What are your recollections about the origins of the Bernoulli Society?

**WRvZ:** The founding of the Bernoulli Society involved a number of groups with different interests. Some of the participants have written about these early days from their own perspective, and when I read some of these accounts, I find it difficult to believe we are talking about the same event.

My primary interest was the European Meeting of Statisticians (EMS). As I said earlier, Jim Durbin and Henri Theil got IMS to start these meetings as IMS Regional Meetings. That was very nice of IMS. I don't think they spent any money on this adventure, but they did proudly advertise these meetings organized by their European Regional Committee, or ERC. Things went perfectly well, and the meetings even survived the almost-death experience in 1969 and branched out to Eastern Europe in 1972. However, some discontent already surfaced during the EMS in London in 1966. A meeting of the participants was called by the local organizers where some participants—mainly British—voiced concern that we were being colonized by the Americans, as one of them put it. I found it difficult to take this seriously, coming from the nationals of a country ruling an empire on which the sun never used to set. However, the argument that we should be able to do these things ourselves without the help of an American society did have some force, even though I knew that IMS had at no time tried to influence any decision made by the ERC. In fact, a little while later IMS had completely forgotten about their European branch and there is no mention of the ERC in their annual report anymore. So, there was some discontent in the air and one of our British colleagues was even hinting darkly at ties between IMS and the US defense establishment.

The next thing that happened was the collapse of the EMS at Hannover in 1969 and we had more important matters to take care of than the relation with IMS. However, after we'd put the EMS back on its feet at Budapest in 1972, I was worried that another collapse of the EMS could happen at any time as long as there was no strong organization behind it. Just working with an *ad hoc* committee without charge or responsibility to report to a higher authority seemed to me an unstable situation. Obviously, IMS couldn't fulfill this role at a distance and we needed a new umbrella for the EMS. So, I thought I'd better start talking to some other people and invited Jim Durbin, John Kingman, Jef Teugels, and Ole Barndorff-Nielsen to meet with me during the 1973 Lunteren meeting. As I had been elected to membership of ISI in 1971, and had developed a perfect working relationship with ISI director Bart Lunenberg, I also spoke to Bart. Earlier ISI had not been interested in the EMS and left it to IMS. Now, Bart wanted to open up ISI beyond its small circle of elected members and thought that sections of ISI that anyone could join would be a good vehicle for that. In ISI, there was a section called the International Association for Statistics in the Physical Sciences (IASPS). Incidentally, starting IASPS had originally been an idea of Jerzy Neyman, who had wanted to create a counterweight to IMS and call it Bernoulli Society. The longish name of IASPS was caused by the fact that Neyman had wanted to exclude the social sciences. The president of IASPS was now David Kendall. At Lunteren, we quickly came to the conclusion that if we could somehow broaden this ISI section to include all of mathematical statistics and probability, it would be a natural home for the EMS.

For various reasons, it seemed best for me not to have anything to do with any discussions with IASPS. As Jim Durbin and Joe Gani had every opportunity to discuss matters with David in Britain, there was no role for me anyway.

While this was going on, a third party entered the discussion—the group organizing the meetings on Stochastic Processes. This was an initiative of Julian Keilson and N. U. Prabhu, and Jef Teugels was very active in this group. They also felt the need for an umbrella organization, and in 1975, IASPS was turned into the Bernoulli Society for Mathematical Statistics and Probability, a section of ISI with three committees responsible for the already ongoing different activities: EMS, the Stochastic Processes Meetings, and meetings on statistics in the physical sciences. David Blackwell became the next president, Jim Durbin treasurer, Jef Teugels scientific secretary, and I chaired the European Regional Committee. I wrote to C. R. Rao who was president of IMS to explain that we were grateful to IMS for starting the European Meetings, but that our committee would now continue as a committee of the new Bernoulli Society. I felt a little bad about this, but needn't have worried. By now, IMS had completely forgotten that it ever started something in Europe, and Rao was obviously mystified by my letter. He wrote back that he hoped that I would advise the members of this new society to attend the upcoming ISI Session in India.

There was one warning voice at the time. Bart and I visited Maurice Kendall, past treasurer of ISI and responsible for the World Fertility Survey that ISI was involved in, and asked his advice. Maurice thought these sections were a pretty good idea, but warned that they might become the tails wagging the dog. Thirty years later, I think ISI will have to come to grips with this problem! In fact, Nick Fisher's ISI committee recently sought to define the role of ISI over and above the activities of the sections that cater to the different branches of statistics, such as mathematical statistics, official statistics, sample survey statistics, statistical computing, biostatistics, statistics in business and industry, and the teaching of statistics.

**Interviewer:** We'll get to ISI in more detail after the next round of drinks, but let's follow the adventures of the Bernoulli Society for a moment.

**WRvZ:** Apart from the already existing activities of the three committees, the Bernoulli Society developed two new activities: the World Meetings and the journal Bernoulli. I would like to talk a little about the first main event which was the World Meeting in Tashkent in 1986. This was a unique occurrence and it took 11 years to bring it off.

At the ISI Session in Warsaw in 1975, I first met Albert Shiryaev. At that time, his English was only slightly better than my Russian, so we must have been conversing partly in sign language. I told him that we had this ERC committee organizing the EMS meetings and that we had developed good contacts in Eastern Europe. When we invited Russian speakers to the EMS, we never had any success. Would it be possible to do a little better in this respect and would it help to have a Russian member on the ERC? He answered yes to both questions. He explained that there were people who would never get permission to come, but other people would sometimes have a better chance. I made it clear that we were only interested in serious scientists rather than people identified with the regime, and that was obvious to him, too. So, he joined the ERC and for every EMS he gave us some names of people likely to turn up if invited, presumably after having discussed this in Moscow. This worked beautifully and we had some excellent people at the EMS. Again, there was, of course, the moral question whether you should invite some people if it is impossible to get some others to come, but we decided to go ahead with this. We figured that by inviting people who couldn't possibly come we would not be doing anybody much good.

If I may digress for a moment, a similar discussion about the best way to deal with this problem killed the Berkeley Symposium. After skipping 1975, the next symposium was planned for 1980. There was a certain amount of battle fatigue among the Berkeley faculty, but Stanford had agreed to take part in this undertaking. The problem centered on whom to invite from the Soviet Union. Neyman felt that the Soviet Academy should be asked to send a delegation and everyone they named should be invited. The other faction felt that the organizers should invite a group of people, and if one of them couldn't come, the invitation for the entire group would be withdrawn. To me, both strategies seemed useless. Even though the Soviet Academy generally behaved quite well, they wouldn't be able to avoid including some less desirable people in the delegation. On the other hand, inviting people who wouldn't possibly get permission to come and then canceling the invitation of all of the others, seemed rather senseless, too. I tried to explain to both groups that the policy of the European Meeting of getting advice from some trusted Russian colleagues was to be preferred, but nobody would listen. That was the end of the celebrated Berkeley Symposium.

**Interviewer:** You were talking about the Tashkent meeting and the Bernoulli Society…

**WRvZ:** Yes, having secured Soviet participation at the European Meetings, Albert and I started thinking how we could have a European Meeting in the Soviet Union. As I said, it took 11 years to get the meeting at Tashkent agreed by one and all. Of course, it would be a bit unusual to have a European Meeting in Uzbekistan, so we invented the concept of a World Meeting of the Bernoulli Society. As I said earlier, it was a great meeting and a wonderful opportunity to meet people and see the world.

**Interviewer:** What about the relationship of IMS and BS in general? You've been President of both…

**WRvZ:** The purposes of IMS and BS are almost identical, so they should cooperate as much as possible. Journals are definitely an area where IMS and BS should collaborate to increase the subscription base and I'm happy to see that this is now happening. Joint meetings have also been successful.

Purely logically, there is really no reason for having two clubs doing the same thing. Peter Bickel has also been president of both IMS and BS, and at one point we asked the IMS council, "Why not merge the two societies?" Of course, we got our heads chewed off. I guess the point is that the leadership of both societies is pretty internationally minded, but on a practical level the membership does have a lot of legitimate local concerns that would be hard to serve from a single centre in the world. It would really need a lot of thought. We never raised this point in BS, but I'm sure the reaction would be the same. Steve Stigler added the argument that competition is a good thing, but I'm not sure this holds for scientific societies as much as for manufacturing TV sets.

**Interviewer:** You have had a long association with the ISI stretching back to… the Paris Session in 1961?

**WRvZ:** Yes, that was the first scientific meeting I attended, but after that I preferred to go to the European Meetings of Statisticians. In those days, ISI Sessions were still predominantly devoted to official statistics. However, in 1971, I was elected a member of ISI. This was supposedly a big thing, with the Dutch members first deciding whom they'd nominate, and then the entire membership voting. Jim Durbin once told me that when he was elected, he was congratulated not just on being elected but on being elected on the first try. I've attended most of the Sessions since then.

**Interviewer:** Do we really need ISI, or could we just as well do without it?

**WRvZ:** We need ISI because it is the only possibility to keep our profession together and keep talking to each other as well as to the outside world. I had never met the director of Statistics Netherlands before I started organizing the ISI Session in Amsterdam, and I doubt that he knew many mathematical statisticians personally. All branches of statistical science have something to contribute to society, but we can only influence matters if we combine forces. Through the various ISI Sections, we possess all of the specialized knowledge we need, our broad constituency allows us to tie things together and our relations with public decision makers are many. ISI can and should speak out on matters of public interest and be heard. So, my answer is that we need ISI, but we should do more with ISI than we have in the past.

**Interviewer:** Have some of the Sessions you've attended been particularly memorable?

**WRvZ:** I'll never forget the Centenary Session in 1985 at Amsterdam because I was the organizer. I messed things up a bit during the opening ceremony. Queen Beatrix of the Netherlands was attending the opening ceremony. In fact, she was probably the only one present who could say that her grandfather attended another ISI Session, which was the one in The Hague in 1913. After the members of the Executive had been introduced to her, I marched into the auditorium with the Queen and we sat down in the front row. I had to make some opening remarks and someone from Statistics Netherlands would put my notes on the lectern so that I wouldn't have to fumble in my pockets. So, I walk up the podium and to the lectern, and, of course, there are no notes. This was a little hard because you have to thank all sorts of people in the right order. Finally, I saw the guy with the notes, but by that time part of the audience was laughing their heads off. When I'd finished and sat down again, the Queen said, "That wasn't very smart was it."

We had this wonderful reception in the Rijksmuseum, where you have the *Night Watch* and all the other famous Rembrandts, so with a glass in your hand you can walk between the great paintings. However, I hadn't realized that when there is a big international reception, all of the pickpockets in Amsterdam assemble at the exit by the time the reception is over. So, quite a few people were robbed and lost cameras and wallets. The next morning, the queue at the police station was really quite something. These were the less successful moments of the Centenary Session, but everything else went well.

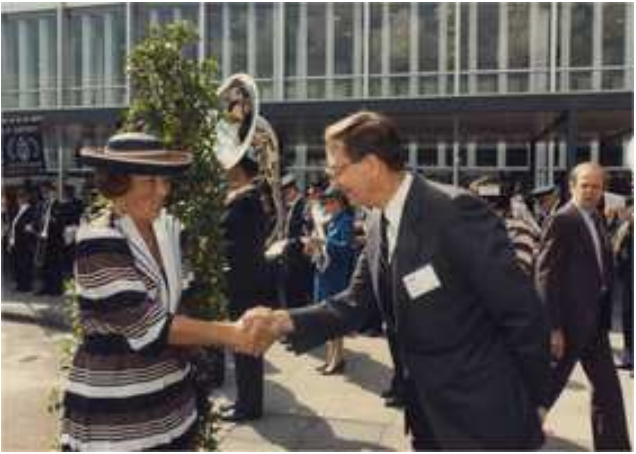

Fig. 4. *Bill greeting Queen Beatrix of the Netherlands arriving for the Opening Ceremony of the Centenary Session of ISI at Amsterdam in 1985.*

Let me just mention another meeting with royalty, at the Session at Tokyo in 1987, where we were presented to Crown Prince Akahito, the present emperor, and his wife. Half an hour in advance of this event, the Japanese organizers put us in a room where they had drawn circles on the floor where we were supposed to stand, and then put a ribbon on everybody's coat. I presume the ribbon said in Japanese who you were and were you came from. Of course, you can't get five statisticians to stand still for half an hour so every once in a while the organizers had to put us back into our circles. When the Crown Prince came along, he talked to Fred Mosteller and said, "I see you are from Harvard University. I am going to visit your beautiful country in a few months and I will meet with your President." Apparently, Fred's thoughts were somewhere else and he figured the Crown Prince was talking about the President of Harvard. So, he replied, "Oh, Dr. So-and-so." You could see the Crown Prince thinking this one over and looking a bit bewildered, he said, "No, no, Mr. Reagan."

To me, he said, "I hear your Queen is very ill," which was true. She was in hospital with pneumonia.

He added, "Next time you see her would you please give her my regards," and I said, "Of course, your Highness" without mentioning that this might take quite some time.

Fred Mosteller was a great man, and a kind person with a fine sense of humor. He didn't put up a show of force, but people listened when he spoke. We spent two years together on the ISI Executive and we were always on the same side in any discussion. His mind was always busy with something, but not necessarily in the here and now. He used to call me up from the US to discuss something, but after dialing all these intercontinental digits, he started thinking of something else while the phone was ringing. When one of my sons picked up the phone, he wouldn't hear anything on the other side. After some time, they got used to this and just said, "Yes, Professor Mosteller."

## 5. OBERWOLFACH

**Interviewer:** You were a frequent invited visitor to the Mathematical Research Institute in Oberwolfach. Would you mind telling us what Oberwolfach is about?

**WRvZ:** Well, Oberwolfach is my favorite mathematical hiding place on earth. Imagine a small village among the hills of the Black Forest in Germany where you suddenly find the Mathematical Research Institute Oberwolfach. Here, they organize one-week conferences about every conceivable topic in mathematics during 51 weeks in the year. Participation is by invitation only. They pay for your stay, but travel is your problem. The Institute consists of a group of modern buildings. There is a building with comfortable rooms for about 40 guests, as well as a dining room, and a place to sit with a self-service bar. They put your napkin in an envelope with your name on it, and at mealtimes they randomly distribute these envelopes over the 6-person dinner tables. This ensures that you really get in touch with everybody else. The quality of the food has varied over the years, but recently it's been excellent. A second building contains the main lecture room, a number of smaller meeting rooms, lots of space to sit around, a music room, billiard tables, table tennis, and last but not least, the best mathematics

---

> When Friedrich Pukelsheim was showing his wife around [Oberwolfach] and saw me sitting in one of the rooms, he said, "... and this is Professor van Zwet, who is part of the furniture."

library in the world. The program of the meetings includes spare time after lunch until 4 p.m., which gives people an opportunity to go hiking in the hills. More recently, they also provide an opportunity for longer research stays for small groups of two or three people. This program is called "Research in Pairs."

The Institute started in 1944, officially as part of the German war effort. Alternatively, it has been viewed as a refuge for people from Freiburg and other nearby universities against the bombing near the end of the war. Recent findings seem to suggest that this early history is a little shady. After the war, the place was financed by the Volkswagen Foundation, and more recently by the state of Baden-Würtenberg and the Federal Government. I love the place and I've been there forty times which must be something of a record for a non-German. When Friedrich Pukelsheim was showing his wife around the place and saw me sitting in one of the rooms, he said "... and this is Professor van Zwet, who is part of the furniture."

**Interviewer:** The place must have changed quite a bit over time.

**WRvZ:** When I first came there in 1969, the building with the guestrooms was brand new, and the lectures were still in the original old mansion. The vast majority of the participants were German and almost all lectures were in German, too. There were only two or three foreign participants. The atmosphere was pretty formal and at dinner, you had the feeling that you were not supposed to say anything until addressed by the senior German professor at your table.

As the number of foreign participants increased, the atmosphere changed noticeably. There were more lectures in English, and I still remember the day when the first German participant spoke in English. It was Fred Eicker from Dortmund, who started his talk in German, explaining to his colleagues that he had been asked by some foreign participants to speak in English, and then proceeded to do so. I think his senior colleagues didn't like this much and some had difficulty understanding English. In the evenings, the international participants—many of whom didn't speak German—used to sit together with a bottle of wine, and after a while a number of the younger Germans joined this international table. So, after a few years, we fit in pretty well.

**Interviewer**: How did Oberwolfach acquire such a great reputation worldwide?

**WRvZ:** Well, Germany has always been important in mathematics and mathematics is important in Germany. Even though this institute started more or less by accident, they really invested in it in a big way over the years. Also, they were the first to develop the concept of an institute with only short time visitors in a secluded place. It has now been copied in France, the US and Canada

**Interviewer:** Do you recall the Oberwolfach Workshop on Robustness that involved John Tukey? Frank Hampel and Helmut Rieder were co-organizers.

**WRvZ:** Yes, it was a strange meeting. We had some evening sessions, which was against all tradition. We were supposed to be finished by six and then you had the rest of the night to do whatever you pleased, but some people started to organize things in the evening. Tukey had this problem-session thing, where he handed out problems to everybody. Rudy and I didn't approve of such activities at night and went to have a drink at Hirschen, the local pub.

We showed up later and found that this evening session was still going on. When Tukey came up with the next problem, I made a serious mistake by saying, "I think I know what to do about that," but that was the wrong answer. He said, "I am not asking you what you think, but what you will do."

By the way, have I ever told you Galen Shorack's story about Tukey? Galen wanted to speak to Tukey about something and made an appointment. When he showed up, it was a hot day and Tukey was sitting outside wearing shorts. As they were talking, Galen notices a wasp circling around and finally landing on Tukey's right knee. Galen was about to warn him, when Tukey suddenly slapped his left knee and got the wasp. Galen said, "This is when I decided that John Tukey was a great statistician."

Pfanzagl and Witting, the two major forces in German statistics, jointly organized an Oberwolfach Workshop on asymptotic statistics. Pfanzagl was interested in second order asymptotics and he opened the meeting by saying how happy he was that all these people doing second order things, like Peter Bickel, Dimitri Chibisov and I were present. He more or less forgot to mention the great majority of other people in the audience, some of who were pretty outstanding. That set the tone for the meeting, which produced some pretty fierce debates. Things got a little confused, and halfway through the meeting, Witting felt he had to speak in defense of applied statistics, which was normally not his favorite topic.

At one point, Pfanzagl himself gave a talk and the program said he would speak for 20 minutes. After that, his collaborator Wolfgang Wefelmeyer would give a related talk for 50 minutes. When Pfanzagl had spoken for half an hour, Chibisov, who was chairing, got up and stood there looking at Pfanzagl, who didn't budge. Chibisov sat down. But then he got restless again, after forty-five minutes, and pointed out to Pfanzagl that he was supposed to speak for only twenty minutes. Pfanzagl said this was not a problem because Mr. Wefelmeyer would just shorten his talk. Poor Wefelmeyer, who was probably there for the first time with a carefully prepared 50-minute talk, just had to cope.

Very late that evening there was some whispering and quiet laughter at a neighboring table from where I was sitting. I went over and found two young people, Friedrich Götze and Christian Hipp, preparing a mock program for the next day. It started with a 90 minute lecture by J. Pfanzagl entitled, "On optimal stopping," continued with a 2-minute talk by W. Wefelmeyer, and went on like that. They carefully removed the official program for the next day from the wall where it was always displayed and replaced it with their own. Realizing that I was witnessing a historic event, I offered to buy the mock program for one bottle of wine. They agreed, on condition that I would remove the program myself after everybody had seen it the next day. This was not without risk, because if anybody would see me, they would certainly decide that I had carried out this prank. However, we agreed and the next day I got hold of the program without being spotted. In the next few years, I sold quite a few Xerox copies for the price of one bottle of wine each. How's that for business instinct!

**Interviewer:** There's an apocryphal story about your April Fools' lecture at Oberwolfach...

**WRvZ:** In March 1976, I was driving to Oberwolfach with John Kingman and Dick Dudley as passengers. For some reason, they happened to be in the Netherlands just before the Oberwolfach meeting. At one point, John says, "Wednesday is April 1 and we really should do something about that." So, we made a pact that whoever spoke on Wednesday would give an April Fools' lecture. What I didn't know, of course, is that when we arrived, John went to the organizer Peter Gänssler and said, "Bill should speak at nine o'clock on Wednesday," without giving a reason for this unusual request. On Tuesday evening, they put up the program for Wednesday and to my great surprise I saw that I was speaking at nine o'clock. I didn't suspect foul play, but of course I should have.

First of all, I thought I had better not take too much of a risk, because they may not like this in a serious place that Oberwolfach still is. So, I went to Gänssler and I said, "Peter, the following is going on, and if I give an April Fools' lecture will I be shot and buried?" He said, "This is okay if you give a serious lecture after the April Fools' one." So, I said fine. Then I had another problem: What if the audience just sits there and keeps writing it all down? I spoke to John, who said, "If you really get in trouble, and nobody reacts, then I will ask a question that will make it clear that it is all nonsense."

So, I start my lecture and introduce a Chinese mathematician by the name of Li who has done all sorts of wonderful things. As I go on, Li's results get crazier and crazier, but nobody reacts, everybody sits there writing it all down. At one point, I stand right next to Georg Neuhaus' chair and say, "Georg, this means that not only Le Cam's first lemma is false, but also his second and third." No reaction. I look imploringly at John who just sits there laughing. In the end, all I can think of is to write April 1 on the blackboard and there is a lot of applause. During the weekly hike on Wednesday afternoon, one by one, people walked beside me for a stretch and said that, of course, they knew it was all nonsense, but didn't want to spoil the fun.

Peter Bickel, Friedrich Goetze, and I spent some time working in Oberwolfach even before the "Research in Pairs" program existed. Now they have formalized it and you can apply for it. It is really great. Just working and hiking a bit. On the weekend, you don't get dinner, but just down the hill is Hirschen, where they serve excellent food.

## 6. EURANDOM

**Interviewer:** You have been associated with many different institutions. In particular, you were the driving force in founding EURANDOM. Where did EURANDOM come from?

**WRvZ:** The name itself is an attempt to express that this was to be a European institute for the study of randomness. Formally, it is an acronym for European Unit for Research and Analysis of Non-Deterministic Operational Models, but this is a well-kept secret. I viewed it as a provisional name until something better occurred to us, but at some point, John Kingman said, "Why don't you keep it, it doesn't sound so bad." So, we did.

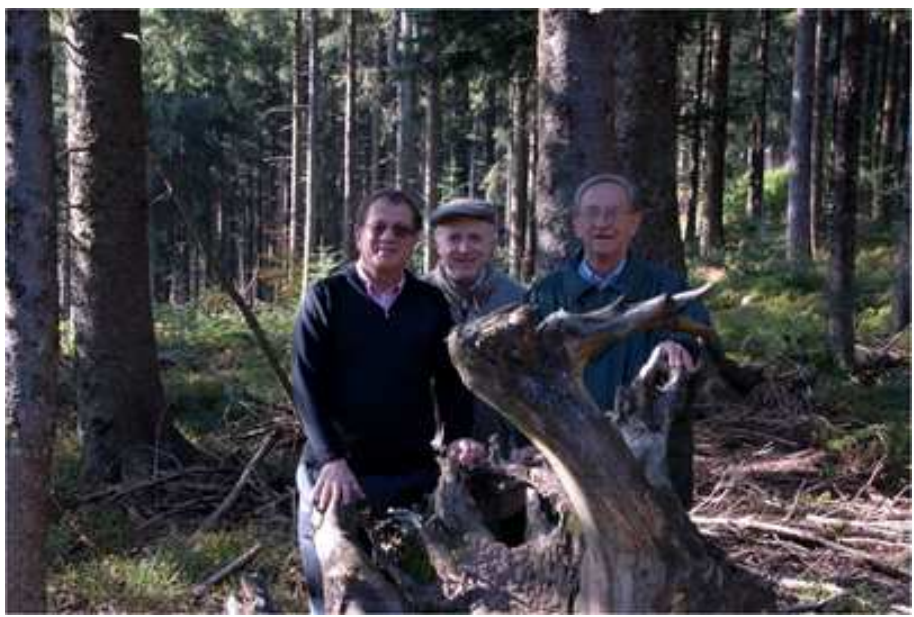

Fig. 5. *From left: Peter Bickel, Friedrich Götze and Bill during a 'Research in Pairs' stay at Oberwolfach in 2007.*

What happened was one of these unusual moments when an opportunity suddenly turns up. An uncle of mine used to say, "When Dame Fortune passes by, you have to grab her by her hair at once, because the back of her head is bald."

In January 1988, some mathematicians including me had a conversation with the Director General for Higher Education of the Dutch Ministry of Education. He explained that compared to fields like physics, mathematics was almost invisible at the Ministry. He suggested that the Minister should appoint a committee to report to him on the state of mathematics in the Netherlands. If such a report would make any sense at all and would also contain some sensible plans, then the Minister might actually fund one or two of these. This sounded like a good idea. Of course every Dutch mathematician wanted his friends on this committee, but after a good deal of infighting, the Minister finally appointed the committee in early 1989. It had an eminently sensible senior analyst as chair, a geometer, an OR person, an engineer and a physicist as members and—somewhat to my surprise—me as vicechair. My surprise was caused by the fact that I knew that my own department chair had been campaigning against my membership of this august club, because he would have much preferred yet another pure mathematician. In his defense, I should add that he later told me that I hadn't behaved too badly.

Anyhow, the committee worked for three years through 1989–1991 and came up with a solid report with sensible proposals to improve the quality of mathematics in the country and increase collaboration of Dutch mathematicians. However, I felt that we should also stress the international angle. Compared to the US with its many large statistics departments, statistics research in Europe was mostly a small scale affair conducted by 3 or 4 people in a math department. To attack larger statistical problems, European collaboration seemed indicated. Because the statisticians had excellent relations all over Europe, starting a European statistical institute seemed an interesting idea. I wrote a large portion of the report anyhow, so one day I inserted this institute into the draft report. My fellow committee members immediately woke up and changed the word "statistics" to "geometry" or "number theory," or whatever. The next time I rewrote this section, I changed this back to "statistics," so the oddnumbered versions of the report recommended this European statistics institute, whereas the even-numbered ones preferred something else. In the end, the chairman felt that in view of the importance of statistics for society, "statistics" would make most sense, although other possibilities should not be excluded. The final version of the report recommended something like "A European institute in a branch of mathematics, for instance statistics." This made everybody happy, and I realized that nobody at the Ministry would misunderstand the message.

So, in February 1992, we handed our report to the Minister. I was in Chapel Hill at the time, but came over to see how the report would be received. The Ministry people seemed quite positive. In fact, a little later the Minister adopted the report, which means he was not opposed to it.

I had extended the activities of the proposed institute to probabilistic OR and probability, and got my colleagues Jaap Wessels and Mike Keane to support this plan Now there were three of us and we all felt it was now a broader and more attractive proposal. However, at that moment, I had other things to do and let it slide. But near the end of the year, I got a phone call from someone I knew at the Ministry, who told me they were preparing next year's budget, and should they put in some money for this institute under a budget item named "internationalization?" There was actually real money under this heading, but if I wanted part of this, I'd better hurry up with a proposal. So, I wrote a 20 page proposal, and after a lot of discussion back and forth with the Ministry people, everybody was happy and the final version formally went to the Minister.

**Interviewer:** That was it?

**WRvZ:** Not quite. Out of the blue, there was a phone call from someone at the Netherlands National Science Foundation (NWO) who said, "Why do you send a proposal like that to the Ministry without consulting us first? We are supposed to advise the Minister about this and we can't support you if you don't consult us first." So, I wrote a letter to NWO saying, "here is a copy of a proposal we sent to the Ministry. We would be very grateful for any support you could give us." In return, I received a letter saying they voted one million guilders for the preparation of this project. I couldn't believe my eyes so I called them up and said, "Where's the cheque?" They said, "Well there is a cheque, but you can only wave it about, it is not a real cheque, there is no money yet, but you can tell everybody else you got a million from us." But in return, they would like us to show that our European colleagues supported us. So, we got together twenty of the leading people in probability, statistics, and OR from all over Europe, and they all agreed this was a pretty good plan. Someone from the Ministry and someone from NWO were sitting there and they were impressed. They were all familiar with Sir John Kingman as a former head of the Science Research Council in Britain, so even the simple fact that John and I knew each other clearly worked in our favor. So, things looked very promising.

A little later things didn't look so good. They told us that, unfortunately, the money had gone. The Minister had used it for something else. However, it turned out this was no problem, and in the end, the Ministry gave us ten million guilders. Then there was a committee, chaired by a former Cabinet Minister and with John Kingman on it once more, that decided that EURANDOM should be at Eindhoven, and the University of Eindhoven chipped in the rest of the money to fund the whole thing for five years. EURANDOM got off the ground in September 1997.

**Interviewer:** What was your role initially at EURANDOM?

**WRvZ:** NWO had cast me in the role of scientific director for the two years remaining until my retirement at Leiden. They insisted and Leiden lent 50% of my time to EURANDOM. It was a busy time and I was actually there a bit longer, but then Frank den Hollander and Henry Wynn took over as joint scientific directors for the next five years.

**Interviewer:** What exactly is EURANDOM doing?

**WRvZ:** Research in stochastics of every description—theoretical as well as applied—by postdocs from all over Europe (and elsewhere occasionally) on a two-year appointment. In my time, there were about 25 postdocs and a small number of Ph.D. students. Guidance is provided by senior people, mostly from various Dutch or Belgian universities, who show up one day a week or so. There are regular visitors from abroad as well as a EURANDOM professor, who is elected every year. The postdocs are an enthusiastic group, who really enjoy contacts with their own age-group and often keep in touch long after they have left. Some are from countries where there a presently few academic jobs and for them this is a chance of a lifetime. Some 20 former EURANDOM postdocs now hold permanent positions in the Netherlands.

**Interviewer:** Is funding ensured for the future?

**WRvZ:** Well, this is always a permanent worry for institutes like this. All I would say is that EURANDOM has been in business for 10 years now and recently had a really superb review from an international visiting committee. It would be the worst possible public policy to stop funding it now.

## 7. FINAL ORDERS

**Interviewer:** If you were starting out again, knowing what you know now, would you do it again? Do you have any idea what field or what research problems you might be interested in?

**WRvZ:** I think I don't have any reason not to pick statistics again.

**Interviewer:** Despite all the changes?

**WRvZ:** Despite all of the changes, statistics and probability is really my home. Whether I'd want to work in a university, I really don't know.

I never thought about any alternative to the university, but I think I might today. You see the trouble is that quite apart from the fact that we seem to be educating large numbers of morons, we have really become small businessmen spending a lot of time looking for money. Over the years, I think I collected more money in grants, etc. than the university ever paid me. Well, if this is our job, then maybe we should do this job elsewhere and get paid for what we do. Also, it is not a very dignified way to earn a living. You are appointed because you are supposedly good at research that makes some sense, so why do you have to spend your days begging for money to carry out this research? And be turned down by some committee that figures that they already spent enough money on something silly like the mathematical sciences.

However, let me make it perfectly clear that I'm not blaming the university for this. It is the way in which public universities are funded that is doing this to us. My farewell lecture at Leiden was entitled "No complaints so far" and I haven't discovered any major complaints since.

**Interviewer:** What is now expected of professors?

**WRvZ:** You are supposed to do three things, Administrative jobs, teaching. and research, and you are only educated for the last of these. Of course,

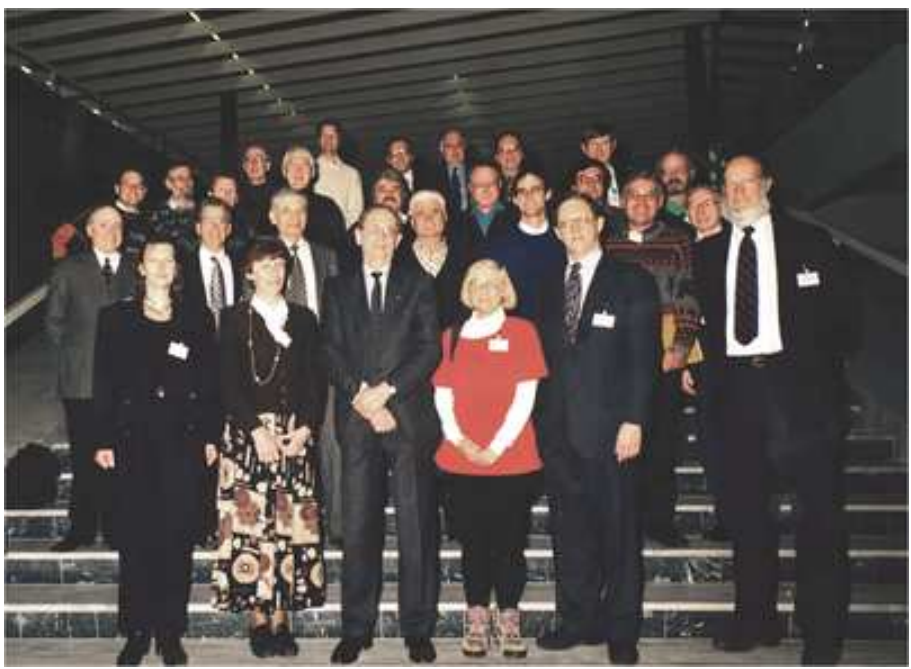

FIG. 6. *Many of the participants of the Symposium for Bill's 65th birthday in Leiden in 1999 can be seen in this picture. The people in this picture as identified by Steve Stigler:*

```
                    22      23     24   25   26    27
        13   14  15    16      17       18      19   20   21
        7        8         9       10            11      12
            1         2         3         4          5          6
```

1. Mathisca de Gunst
2. Marie Hušková
3. Bill van Zwet
4. Grace Wahba
5. Steve Stigler
6. David Siegmund
7. Friedrich Götze
8. Chris Klaassen
9. Dimitri Chibisov
10. Rafail Hasminskii
11. Adrian Baddeley
12. Peter Hall
13. Richard Gill
14. Hermann Dinges
15. Claude Belisle
16. Ole Barndorff-Nielsen
17. Sándor Csörgő
18. David Mason
19. Lucien Birgé
20. Jon Wellner
21. Ron Pyke
22. Peter Jagers
23. Aad van der Vaart
24. Ildar Ibragimov
25. Fred Bakker
26. Rudy Beran
27. Iain Johnstone

a certain amount of administrative work is unavoidable, unless you want to fill the university with managers without any affinity to science. The problem is that many people escape from doing their bit because of real or imagined incompetence for such things. So, other people have to do more than their share. For a couple of years, I served as Dean of the School of Sciences, which was a full-time job. Once you are in this administrative environment, it takes determination to get out. I was pressed to become Rector of the university twice. If you do that, there is no return and your scientific career is over. The first time this happened was in 1984 when I was fifty and not ready to quit statistics. So, I was firm. It is somewhat ironic that instead of running the university, I choose to run *The Annals of Statistics*, which is almost as bad, but at least you learn something. It took 80% of my time for $3\frac{1}{2}$ years, but it was fun working with a really superb editorial board.

When all is said and done, there are still advantages to university life. I mean your freedom is considerable.

**Interviewer:** What have you worked on that you have really enjoyed?

**WRvZ:** Well the things that I always liked best are problems that are really complex, until you look at them in the right way and there is a simple and elegant solution. I really love that. That's probably the Hoeffding side of my character. Most students feel bad about such things because they figure that a problem with a simple solution is trivial and, therefore, no good as a thesis subject. But the best thing that can happen to you is a really complicated problem that by some trick, you can see through. It doesn't happen every day, but it happened to me with my 1978 paper on Kakutani's interval splitting (van Zwet, 1978). At the 1976 Oberwolfach meeting where I gave my April Fools' talk, Dick Dudley told us about a problem that Kakutani had proposed at an analysis meeting six months earlier. You pick a random (that is uniformly distributed) point in the unit interval. Now you have two intervals and you put another random point in the longer of the two. Continue in this way, each time putting a random point in the longest of the intervals that you have at that moment. Show that as the number of points increases, they become uniformly distributed over the unit interval. If you think about it, it must clearly be true, but the obvious attempts at a proof look hopelessly messy. All of us thought about this question throughout the week, but nobody got anywhere. At the end of the week, we awarded first prize to John Kingman who claimed to have a proof that would work if $\pi^2 > 19$ or something like that. Still the solution is elementary and elegant, once you look at it in the right way. After thinking about this on and off for another 25 years, Ron Pyke and I showed that the empirical process of this sampling scheme tends to a constant times the Brownian bridge (Pyke and van Zwet, 2004)—really neat!

I have a similar experience with a question posed by Herbert Robbins concerning the distribution of electrons on a conducting sphere (van Zwet, 1993). Let me also mention a paper (Bickel, Chibisov and van Zwet, 1981) where we provide an almost trivial explanation of the phenomenon that—in a phrase coined by Pfanzagl—"first order efficiency implies second order efficiency."

There are other cases where the importance of the problem, and perhaps your curiosity about the outcome, keeps you going. To study the properties of statistical procedures more precisely than you can with limit theorems, you need asymptotic expansions. After the initial work for rank tests that I already mentioned, Peter Bickel, Friedrich Götze and I later joined by Vidmantas Bentkus wrote a series of papers laying the foundation of a general theory (van Zwet, 1984; Bickel, Goetze and van Zwet, 1986; Bentkus, Goetze and van Zwet, 1997). This was extremely technical work, but we felt it had to be done. In the end, we got reasonably close to a general theory of these things, but there always remains more work to be done. Some of this also turned out to be of interest when we later worked on resampling (e.g., Putter and van Zwet, 1998). Anyhow, there are so many things you get interested in for so many reasons, that it is hard to say what your primary interests are.

**Interviewer:** How would you characterize your style? Some people, to take extreme cases, are problem-solvers, some people are system-builders.

**WRvZ:** It is difficult to say. A minute ago, I would probably have said that I'm a problem solver, but the last things I mentioned should probably be classified as theory building. One thing I'm sure of is that I never want to get stuck in one particular area. The nice thing about statistics and probability is that problems are everywhere. You can't walk in the street without seeing them.

**Interviewer:** If you were stuck on a desert island with a limited choice of reading materials which of your papers, among those available, would you take with you?

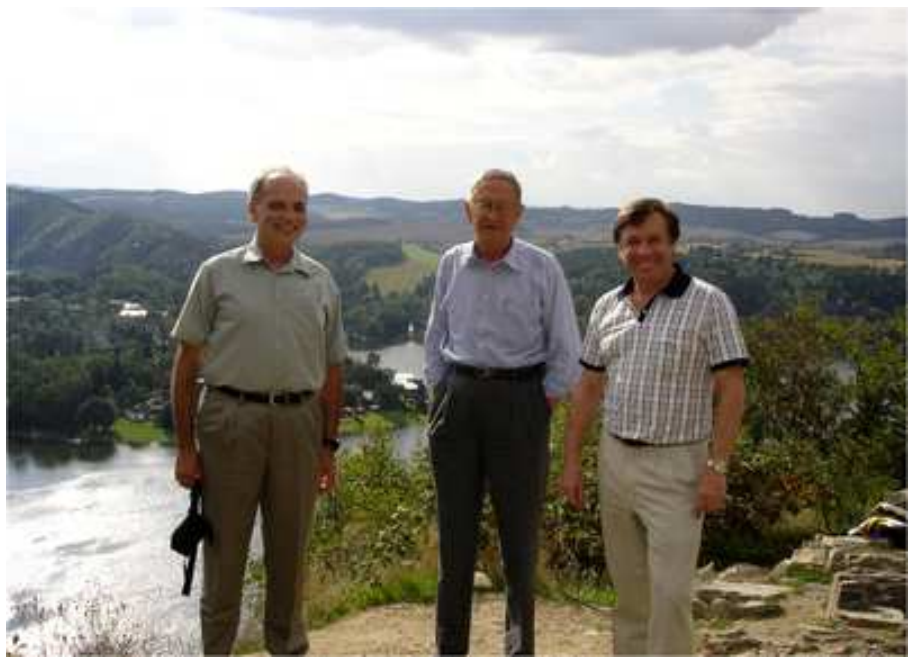

FIG. 7. *From left: Rudy Beran, Bill and Nick Fisher at Slapy Reservoir during Prague Stochastics in 2006, when the interview was recorded.*

**WRvZ:** That is a strange question. Why would I take any of my papers? A crate of Dutch genever or bourbon if you like would make a lot more sense. Or things to sow and grow food, or is that available on your island?

**Interviewer:** Let's say the papers you like best.

**WRvZ:** I really like my Ph.D. thesis (van Zwet, 1964), especially in view of the fact that I knew nothing at the time, so let me bring that along. Two ancient papers with Kobus Oosterhoff on combining tests and contiguity (Oosterhoff and van Zwet, 1967, 1979) and the two Kakutani papers I just mentioned. Some of the asymptotic expansion papers. Two papers with Chris Klaassen and Aad van der Vaart connecting the accuracy of estimating a parameter and its score function (Klaassen and van Zwet, 1985 and Klaassen, van der Vaart and van Zwet, 1988). A pretty paper with my son Erik on a topological investigation of the consistency phenomenon (van Zwet and van Zwet, 1999). A number of bootstrap papers joint with Hein Putter, Peter Bickel, and Friedrich Götze (e.g., Putter and van Zwet, 1996; Bickel, Goetze and van Zwet, 1997) and two papers on plant cell division joint with Mathisca de Gunst (de Gunst and van Zwet, 1992, 1993) and on statistics for the contact process, joint with Marta Fiocco (Fiocco and van Zwet, 2003a, 2003b). Finally, a neat little paper with Nelly Litvak on the time needed to visit $n$ random points on a circle that turns out to be connected to splines and Jacobi's theta functions (Litvak and van Zwet, 2004). That should be enough to fill a sailor's chest. It is also a fair percentage of my total output, which is only about 80-some papers.

**Interviewer:** Was there much pressure on you to publish?

**WRvZ:** When I arrived at Leiden, there was none. If you taught your classes, you were okay. I remember that many years later, someone tried to stop the promotion of a physicist who hadn't published anything for years. His defense was, "When I was appointed nobody said anything about research. I was supposed to teach."

Personally, I set myself a standard of two decent papers a year, and this I kept doing through the years.

What, has the bar closed? That means the evening's over.

## ACKNOWLEDGMENTS

RJB's research was supported in part by National Science Foundation Grant DMS-04-04547. NIF's research was supported by ValueMetrics Australia. We wish to thank the Organizers of *Prague Stochastics 2006*, the conference during which this interview took place, for their generous support and hospitality.

## REFERENCES

Albers, W., Bickel, P. J. and van Zwet, W. R. (1976). Asymptotic expansions for the power of distribution-free tests in the one-sample problem. *Ann. Statist.* **4** 108–156. MR0391373

Barlow, R. E. and van Zwet, W. R. (1970). Asymptotic properties of isotonic estimators for the generalized failure rate function. Part I: Strong consistency. In *Nonparametric Techniques in Statistical Inference* (M. L. Puri, ed.) 159–173. Cambridge Univ. Press. MR0275607

Bentkus, V., Goetze, F. and van Zwet, W. R. (1997). An Edgeworth expansion for symmetric statistics. *Ann. Statist.* **25** 851–896. MR1439326

Bickel, P. J., Chibisov, D. M. and van Zwet, W. R. (1981). On efficiency of first and second order. *Internat. Statist. Rev.* **49** 169–175. MR0633598

Bickel, P. J., Goetze, F. and van Zwet, W. R. (1986). The Edgeworth expansion for $U$-statistics of degree two. *Ann. Statist.* **14** 1463–1484. MR0868312

Bickel, P. J., Goetze, F. and van Zwet, W. R. (1997). Resampling fewer than n observations: Gains, losses, and remedies for losses. *Statist. Sinica* **7** 1–31. MR1441142

Bickel, P. J. and van Zwet, W. R. (1978). Asymptotic expansions for the power of distributionfree tests in the two-sample problem. *Ann. Statist.* **6** 937–1004. MR0499567

Bloemena, A. R. (1964). *Sampling from a Graph. Mathematics Centre Tracts* **2**. Mathematisch Centrum, Amsterdam. MR0175265

Cramér, H. (1946). *Mathematical Methods of Statistics.* Princeton Univ. Press, Princeton. MR0016588

Feller, W. (1950). *An Introduction to Probability Theory and Its Applications. Vol. I.* Wiley, New York. MR0038583

Feller, W. (1966). *An Introduction to Probability Theory and Its Applications. Vol. II.* Wiley, New York. MR0210154

Fiocco, M. and van Zwet, W. R. (2003a). Decaying correlations for the supercritical contact process conditioned on survival. *Bernoulli* **9** 763–781. MR2047685

Fiocco, M. and van Zwet, W. R. (2003b). Parameter estimation for the supercritical contact process. *Bernoulli* **9** 1071–1092. MR2046818

Fisher, N. I. and Sen, P. K. (1994). *The Collected Works of Wassily Hoeffding.* Springer, New York. MR1307621

de Gunst, M. C. M. and van Zwet, W. R. (1992). A non-Markovian model for cell population growth: speed of convergence and central limit theorem. *Stochastic Process. Appl.* **41** 297–324. MR1164182

de Gunst, M. C. M. and van Zwet, W. R. (1993). A non-Markovian model for cell population growth: Tail behavior and duration of the growth process. *Ann. Appl. Probab.* **3** 1112–1144. MR1241037

Hájek, J. and Šidák, Z. (1967). *Theory of Rank Tests.* Academic Press, New York.

Hodges, J. L. and Lehmann, E. L. (1970). Deficiency. *Ann. Math. Statist.* **41** 783–801. MR0272092

Hoeffding, W. (1961). *The Strong Law of Large Numbers for U-statistics. Institute of Statistics Mimeograph Series* **302**. Univ. North Carolina, Chapel Hill.

Kendall, M. G. (1948). *The Advanced Theory of Statistics.* Griffin, London.

Klaassen, C. A. J. and van Zwet, W. R. (1985). On estimating a parameter and its score function. In *Proc. Berkeley Conference in Honor of Jerzy Neyman and Jack Kiefer* **II** (L. M. LeCam and R. A. Olshen, eds.) 827–840. Wadsworth, Monterey. MR0822068

Klaassen, C. A. J., van der Vaart, A. W. and van Zwet, W. R. (1988). On estimating a parameter and its score function, II. In *Statistical Decision Theory and Related Topics* **IV**, **2** (S. S. Gupta and J. O. Berger, eds.) 281–288. Springer, New York. MR0927140

Lehmann, E. L. (1959). *Testing Statistical Hypotheses.* Wiley, New York. MR0107933

Lehmann, E. L. (2008). *Reminiscences of a Statistician. The Company I Kept.* Springer, New York. MR2367933

Loève, M. (1955). *Probability Theory.* Van Nostrand, Toronto.

Litvak, N. and van Zwet, W. R. (2004). On the minimal travel time needed to collect n items on a circle. *Ann. Appl. Probab.* **14** 881–902. MR2052907

Oosterhoff, J. and van Zwet, W. R. (1967). On the combination of independent test statistics. *Ann. Math. Statist.* **38** 659–680. MR0216645

Oosterhoff, J. and van Zwet, W. R. (1979). A note on contiguity and Hellinger distance. In *Contributions to Statistics* (*Jaroslav Hajek Memorial Volume*) (J. Jurečková, ed.) 157–166. Academia, Prague. MR0561267

Putter, H. and van Zwet, W. R. (1996). Resampling: Consistency of substitution estimators. *Ann. Statist.* **24** 2297–2318. MR1425955

Putter, H. and van Zwet, W. R. (1998). Empirical Edgeworth expansions for symmetric statistics. *Ann. Statist.* **26** 1540–1569. MR1647697

Pyke, R. and van Zwet, W. R. (2004). Weak convergence results for the Kakutani interval splitting procedure. *Ann. Probab.* **32** 380–423. MR2040787

Rao, C. R. (1965). *Linear Statistical Inference and Its Applications.* Wiley, New York. MR0221616

Scheffé, H. (1959). *The Analysis of Variance.* Wiley, New York. MR0116429

Schmetterer, L. (1966). *Mathematische Statistik.* Springer, New York. MR0231463

van Dantzig, D. (1957a). Statistical Priesthood I. Savage on personal probabilities. *Statist. Neerlandica* **11** 1–16.

van Dantzig, D. (1957b). Statistical Priesthood II. Sir Ronald on scientific inference. *Statist. Neerlandica* **11** 185–200.

van Zwet, E. W. and van Zwet, W. R. (1999). A remark on consistent estimation. *Math. Methods Statist.* **8** 277–284. MR1722624

van Zwet, W. R. (1964). *Convex Transformations of Random Variables. Mathematics Centre Tracts* **7**. Mathematisch Centrum, Amsterdam. MR0175265

van Zwet, W. R. (1978). A proof of Kakutani's conjecture on random subdivision of longest intervals. *Ann. Probab.* **6** 133–137. MR0478307

van Zwet, W. R. (1984). A Berry–Esseen bound for symmetric statistics. *Z. Wahrsch. Verw. Gebiete* **66** 425–440. MR0751580

van Zwet, W. R. (1993). The asymptotic distribution of point charges on a conducting sphere. In *Statistical Decision Theory and Related Topics* **V** (S. S. Gupta and J. O. Berger, eds.) 427–430. Springer, New York. MR1286318

Wilks, S. S. (1962). *Mathematical Statistics.* Wiley, New York. MR0144404

Yule, G. U. and Kendall, M. G. (1950). *An Introduction to the Theory of Statistics.* Griffin, London. MR0035938